\documentclass[a4paper,10pt]{article}
\usepackage{graphicx}
\usepackage{subfigure}
\usepackage{fancybox}
\usepackage{amsmath}
\usepackage{float}
\usepackage{pifont}
\usepackage{fancyhdr}

\begin{document}

\title{\vspace{-4cm}Three-dimensional calculation of
       inhomogeneous structure in low-density nuclear matter}

\author { Minoru Okamoto$^{1,2}$,
          Toshiki Maruyama$^{2}$,
          Kazuhiro Yabana$^{1,3}$ and
          Toshitaka Tatsumi$^{4}$ }
\date{}
\maketitle

\begin{center}
 $^{1}$Graduate School of Pure and Applied Science, University of Tsukuba,
 Tennoudai 1-1-1, Tsukuba, Ibaraki 305-8571, Japan\\
 $^{2}$Advanced Science Research Center, Japan Atomic Energy Agency,
 Shirakata Shirane 2-4, Tokai, Ibaraki 319-1195, Japan\\
 $^{3}$Center of Computational Sciences, University of Tsukuba, Tennoudai
 1-1-1, Tsukuba, Ibaraki 305-8571, Japan\\
 $^{4}$Department of Physics, Kyoto University, Kyoto 606-8502, Japan
\end{center}
\par

\begin{abstract}
In low-density nuclear matter which is relevant to the crust region
of neutron stars and collapsing stage of supernovae,
non-uniform structures called ``nuclear pasta'' are expected.
So far, most works on nuclear pasta have used the Wigner-Seitz cell approximation 
with anzats about the geometrical structures like droplet, rod, slab and so on.
We  perform fully three-dimensional calculation of non-uniform nuclear matter
 for some cases with fixed proton ratios and in beta-equilibrium based on the
 relativistic mean-field model and the Thomas-Fermi approximation.
In our calculation typical pasta structures are observed.
However, there appears some difference 
in the density region of each pasta structure.
\end{abstract}

\setcounter{page}{1}

\pagestyle{fancy}

\lhead{\slshape \leftmark}
\chead{}

\rhead{\slshape \rightmark}
\cfoot{\thepage}
\renewcommand{\headrulewidth}{1pt}
\renewcommand{\footrulewidth}{1pt}

\pagenumbering{arabic}

\section{Introduction}
In 1934, W.Baade and F.Zwicky proposed the existence of neutron stars
only one year after the discovery of neutron.
Estimating the binding energy of neutron stars, they predicted that
neutron stars are made by supernova explosions.
Neutron star had been an imaginary object for 30 years, until
pulsars were discovered by J.Belle and A.Hewish in 1967.
Neutron star has a radius of about 10 km and the mass about 1.4 times of the sun,
and consists of four parts~\cite{Neutron_star_structure}.
The region around 0.3${\rm km}$ from  outside is called ``outer crust'',
where Fe nuclei are expected to form a Coulomb lattice.
Around $0.3-1{\rm km}$ is called ``inner crust'' with a density about 0.3-0.5$\rho_0$.
There are neutron-rich nuclei in lattice and dripped neutrons in a superfluid state.
Two central regions with higher density are called ``outer core'' and ``inner core'', 
where speculated are
proton super-conductivity, neutron super-fluidity \cite{pi-condensate},
meson condensations \cite{KV03}, hyperon mixture \cite{glenhyp},
or quark states \cite{Glendenning,HQmix}.

The transition of matter composition with the change of density inside neutron stars
causes a question:  does it change continuously or suddenly?
A sudden change of matter property is generally accompanied by a first-order phase transition
which causes an appearance of the mixed phase.

Ravenhall et al \cite{Ravenhall} suggested the existence of
non-uniform structures of nuclear matter, i.e.\ the structured mixed phase.
They suggested five types of structures as droplet, rod, slab, tube, and bubble.
Due to its geometrical shapes which depend on the density, we call it ``nuclear pasta''
like spaghetti and lasagna etc \cite{Ravenhall,Hashimoto}.
Many workers have suggested the existence of pasta structures in
low-density nuclear matter, relevant to the crust region of neutron stars 
and the collapsing stage of supernovae.
The existence of the pasta structures at the crust of neutron stars may not 
influence on  the bulk property and structure of neutron stars.
However, it should be important for the mechanism of glitch,
the cooling process of neutron stars, and the thermal and mechanical properties
of supernova matter.

The pasta structures presented by Ravenhall have geometrical symmetries. 
So we can treat the system  with the Wigner-Seitz (WS) cell approximation.
Because of the convenience, the WS cell approximation has been very often used.
But there may be some shortcomings.
First, the existence of some kinds of structures other than the typical pasta structures were suggested.
For example, 
in compressive process of matter, two droplets connect with each other 
and form dumbbell-like pieces \cite{QMD_1}.
Other examples are 
double diamond~\cite{Fukuoka.Uni,gyroid} and gyroid~\cite{gyroid} shapes of matter
suggested by using compressible liquid-drop model.
These shapes can't exist as a ground state at zero temperature, so they
might exist in supernova matter at finite temperatures.
Such structures are impossible to be described by the WS approximation. 
It is worth trying to calculate without any approximation whether or not
these structures exist as a ground state or an excited state.

\section{Method}
\subsection{Relativistic Mean Field Theory}
There are several approaches for studying pasta structures in the literature, as
liquid-drop model~\cite{Ravenhall,gyroid}, Thomas-Fermi model \cite{Wil85,oyamatsu}, 
Hartree-Fock~\cite{newton},
quantum molecular dynamics (QMD)~\cite{QMD_2,QMD_1}, 
relativistic mean field model (RMF)~\cite{relamaruyama,Toki}.
In the studies using liquid-drop model and RMF model,
always used was the Wigner-Seitz (WS) cell approximation,
where only the typical pasta structures are considered.
QMD calculation does not assume any specific structure of baryons.
But uniform electron distribution, though it should be almost uniform, is assumed.
The Thomas-Fermi calculation in Ref.\ \cite{Wil85} and HF calculation \cite{newton}
used the periodic boundary condition and did not assume any geometrical symmetry 
in structure.
However, the size of the periodic cell was not large enough for quantitative discussion. 

In this paper, we use the Thomas-Fermi approximation for baryons 
with interaction by the RMF model.
This model is not only simple for numerical calculation but also 
quantitatively reliable for the properties of finite nuclei 
and the saturation property of matter.

We start from the simple Lorentz-scalar Lagrangian with nucleons,
electrons and $\sigma$, $\omega$, $\rho$-mesons as follows,
\begin{eqnarray}
\nonumber
L &=&{\bar{\psi}}{\left[i{\gamma}^{\mu}{\partial}_{\mu}
      - m_{N}^{*} - g_{{\omega}N}{\gamma}^{\mu}{\omega}_{\mu}
      - g_{{\rho}N}{\gamma}^{\mu}{\mbox{\boldmath$\tau$}}{\cdot}{\mbox{\boldmath$R$}}_{\mu}
    - e{\frac{1+{\tau}_3}{2}}{\gamma}^{{\mu}} A_{\mu} \right]}{\psi} \\
\nonumber
  &+& {\frac{1}{2}}({\partial}_{\mu}{\sigma})^2  - {\frac{1}{2}}m_{\sigma}^2 {\sigma}^2 - U({\sigma}) 
             -  {\frac{1}{4}}{\omega}_{\mu\nu} {\omega}^{\mu\nu}
             +  {\frac{1}{2}}m_{\omega}^2{\omega}_{\mu}{\omega}^{\mu} 
             -  {\frac{1}{4}}{\mbox{\boldmath$R$}}_{\mu\nu} {\mbox{\boldmath$R$}}^{\mu\nu} 
+ {\frac{1}{2}}m_{\rho}^2 {\mbox{\boldmath$R$}}_{\mu} {\mbox{\boldmath$R$}}^{\mu} \\
  &-& {\frac{1}{4}}F_{\mu\nu} F^{\mu\nu}
     +  {\bar{{\psi}_e}}{\left[i{\gamma}^{\mu}{\partial}_{\mu} - m_e
     + e{\gamma}^{\mu} A_{\mu}\right]}{\psi}_e,
\label{Lagrangian}
\end{eqnarray}
where the effective mass is written as $m_N^* = m_N - g_{{\sigma}N}{\sigma}$, 
and the potential energy of sigma meson $U{\left({\sigma}\right)} = {\frac{1}{3}}bm_{N}{\left(g_{\sigma}{\sigma}\right)^3}
        -{\frac{1}{4}}c{\left(g_{\sigma}{\sigma}\right)^4}$.

From the variational principle for this Lagrangian, we get
 equations of motion for nucleons, $\sigma$, $\omega$, $\rho$-mesons 
and the Coulomb potential as follows:
\begin{equation}
{\left[i{\gamma}^{\mu}{\partial}_{\mu}-m_{N}^{*} 
 -g_{{\omega}N}{\gamma}^{\mu}{\omega}_{\mu} -g_{{\rho}N}{\gamma}^{\mu}{\mbox{\boldmath$\tau$}}{\cdot}{\mbox{\boldmath$R$}}_{\mu}
 -e{\frac{1+{\tau}_{3}}{2}}{\gamma}^{\mu}A_{\mu} \right]}{\psi}=0,
\label{kakushi_eq}
\end{equation}
\begin{equation}
{\partial}_{\mu}{\partial}^{\mu}{\sigma}-{\left({\bar{\psi}}g_{{\sigma}N}{\psi}
 -m_{\sigma}^{2} {\sigma} -{\frac{dU}{d{\sigma}}} \right)}=0
\label{sigma_eq1}
\end{equation}

\begin{equation}
-{\partial}_{\mu}{\omega}^{{\mu}{\nu}}
-\left(-g_{{\omega}N}{\bar{\psi}}{\gamma}^{\nu}{\psi}+m_{\omega}^2{\omega}^{\nu}\right)=0
\label{omega_eq}
\end{equation}
\begin{equation}
\partial_\mu{\mbox{\boldmath$R$}}^{\mu\nu}+m_{\rho}^2{\mbox{\boldmath$R$}}^\nu=g_{{\rho}N}{\bar{\psi}}{\mbox{\boldmath$\tau$}}{\gamma}^{\nu}{\psi},
\label{rho_eq}
\end{equation}
\begin{equation}
{\nabla}^{2}A_0 = e^2\left({\bar{\psi}}{\gamma}^0(1+\tau_3)/2{\psi}-{\bar{\psi}_e}\gamma_0{\psi_e}
		     \right) = e^2{\hat\rho}_{ch}.
\label{Coulomb_eq}
\end{equation}

By the mean-field approximation for meson fields and the static approximation
for electric field,
we have ${\langle}{\sigma}{\rangle}={\sigma}$, 
${\langle}{\omega}^{\mu}{\rangle}={\delta}^{{\mu},0}{\omega}_0$,
${\langle}{R}_a^{\mu}{\rangle}={\delta}^{{\mu},0}\delta_{a,3}{R}_0$
and $V=\langle A_0\rangle$.
The bra-ket $\langle \rangle$ represents the expectation value in the 
ground-state of nuclear matter,
and we assume the space and time-reversal symmetries for this state.
$\omega$ and $\rho$ mesons can't have finite value in space components.
Therefore we finally get the equations for mesons as follow.

\begin{equation}
-{\nabla}^{2}{\sigma}+m^{2}_{\sigma}{\sigma} =
  -\frac{dU}{d\sigma}+g_{{\sigma}B}{\langle}\bar{\psi}{\psi}{\rangle}
\label{sigma_rmf}
\end{equation}
\begin{equation}
-{\nabla}^{2}{\omega}_{0}+m_{\omega}^{2}\omega_{0} =
 g_{{\omega}N}(\rho_{p}+\rho_{n})
\label{omega_rmf}
\end{equation}
\begin{equation}
-{\nabla}^{2}{R}_{0}+m_{\rho}^{2}R_{0} =
 g_{{\rho}N}({\rho}_{p}-{\rho}_{n})
\label{rho_rmf}
\end{equation}
By the Thomas-Fermi approximation at zero temperature,
momentum distributions of nucleons have step-functional forms 
and plane-wave solutions for Dirac equation.
Scalar density and density of nucleons  
${\rho}_s=\sum_{i=1}^{A}\bar{\psi}_i(x){\psi_i(x)}$, 
${\rho}=\sum_{i=1}^{A}{\psi_i(x)}^\dagger{\psi_i(x)}$ 
are written using ${\psi}_i$  ($i=1,2,\cdots A$) as
\begin{eqnarray}
 \nonumber
  {\rho}    &=&
  2\int_0^{k_F}\frac{d^3k}{\left(2{\pi}\right)^3}\frac{m_N^*}{E_k^*}
    \sum_s{u^{\dagger}(k,s)u(k,s)} \\
  &=&
   4\int_{0}^{k_F}\frac{d^3k}{\left(2{\pi}\right)^3}=\frac{2k_F^3}{3{\pi}^2}\\
 \nonumber
 {\rho}_s  &=&   2\int_0^{k_F}\frac{d^3k}{\left(2{\pi}\right)^3}\frac{m_N^*}{E_k^*}
    \sum_s{\bar{u}(k,s)u(k,s)} \\
  &=& 4\int_{0}^{k_F}\frac{d^3k}{\left(2{\pi}\right)^3}\frac{m_N^*}{E_k^*},
\end{eqnarray}
,here we use $E_k^*=\sqrt{m_N^{*2}+k^2}$ and the electron density is given as ${\rho}_{e}\equiv -{\int d^{3}x{\langle}{\bar{\psi}_e}\gamma_0{\psi_e}{\rangle}} <0$.
The energy density of nuclear matter
is obtained by the energy-momentum tensor 
and putting $\left({\mu},{\nu} \right)=\left(0,0\right)$, as
\begin{eqnarray}
 \nonumber
 \hat{T}^{00}_{\rm RMF} &=&
  \langle\psi^{\dagger}\left[-i{\mbox{\boldmath$\alpha$}}{\cdot}{\mbox{\boldmath$\nabla$}}+{\beta}m_{N}^{*}
                 +g_{{\omega}N}{\omega}_0+g_{{\rho}N}R_0
                 +e\frac{1+{\tau}_3}{2}V\right]\psi\rangle\\
 \nonumber & & \hspace{1pt}
  +\frac{1}{2}({\nabla}{\sigma})^2+\frac{1}{2}m_{\sigma}^2+U\left({\sigma}\right)\\
 \nonumber & & \hspace{1pt}
  -\frac{1}{2}({\nabla}{\omega}_0)^2-\frac{1}{2}m_{\omega}^2{\omega}_0^2
  -\frac{1}{2}({\nabla}R_0)^2-\frac{1}{2}m_{\rho}^{2}R_0
  -\frac{1}{2}({\nabla}V)^2 \\
 & & \hspace{1pt}
  +\langle{\psi}_e^{\dagger}
           \left[-i{\mbox{\boldmath$\alpha$}}{\cdot}{\mbox{\boldmath$\nabla$}}
                                                    +{\beta}m_e-eV\right]{\psi}_e\rangle
\label{energy_tensor_RMF}
\end{eqnarray}

Therefore, we get the total energy
\begin{eqnarray}
 \nonumber
 E &=& 2{\sum_{n,p{\in{F}}}{\sqrt{k^2+m_N^{*2}}}} \\
 \nonumber
   & & \hspace{1pt}
   +{\int{d^3x\left[ \frac{1}{2}\left({\nabla}{\sigma}\right)^2
                    +\frac{1}{2}m_{\sigma}^2{\sigma}^2
		    +U\left({\sigma}\right) \right]}} \\
 \nonumber
   & & \hspace{1pt}
   +{\int{d^3x\left[ \frac{1}{2}\left({\nabla}{\omega}_0\right)^2
                    +\frac{1}{2}m_{\omega}^2{\omega}_0^2
                    +\frac{1}{2}\left({\nabla}R_0\right)^2
                    +\frac{1}{2}m_{\rho}^2R_0^{2}\right]}} \\
   & & \hspace{1pt}
   +{\int{d^3x\left[-\frac{1}{2}\left({\nabla}V\right)^2\right]}}+E^{'}_e
\label{Total_Energy}
\end{eqnarray}

\begin{eqnarray}
 \nonumber
 E^{'}_e &=& \int{d^3x \left[\frac{2}{\left(2{\pi}\right)^3}
                   \int_0^{k_F^{\left(e\right)}}{4{\pi}k^2dk\sqrt{k^2+m_e^2}}
                     -eV\left(-\rho_e\right) \right]}\\
 &=& \int{d^3x\left[\frac{{\left({\mu}_e-V\right)}^4}{4{\pi}^2} -V{\rho}_e\right]}
       \hspace{15pt} \left({\mu}_e \approx
		      k_F^{\left(e\right)}+V\right)
\label{Total_Energy_Coulomb}
\end{eqnarray}

\begin{figure}[H]
 \begin{center}
 \subfigure[]{
 \includegraphics[width=0.4\textwidth]{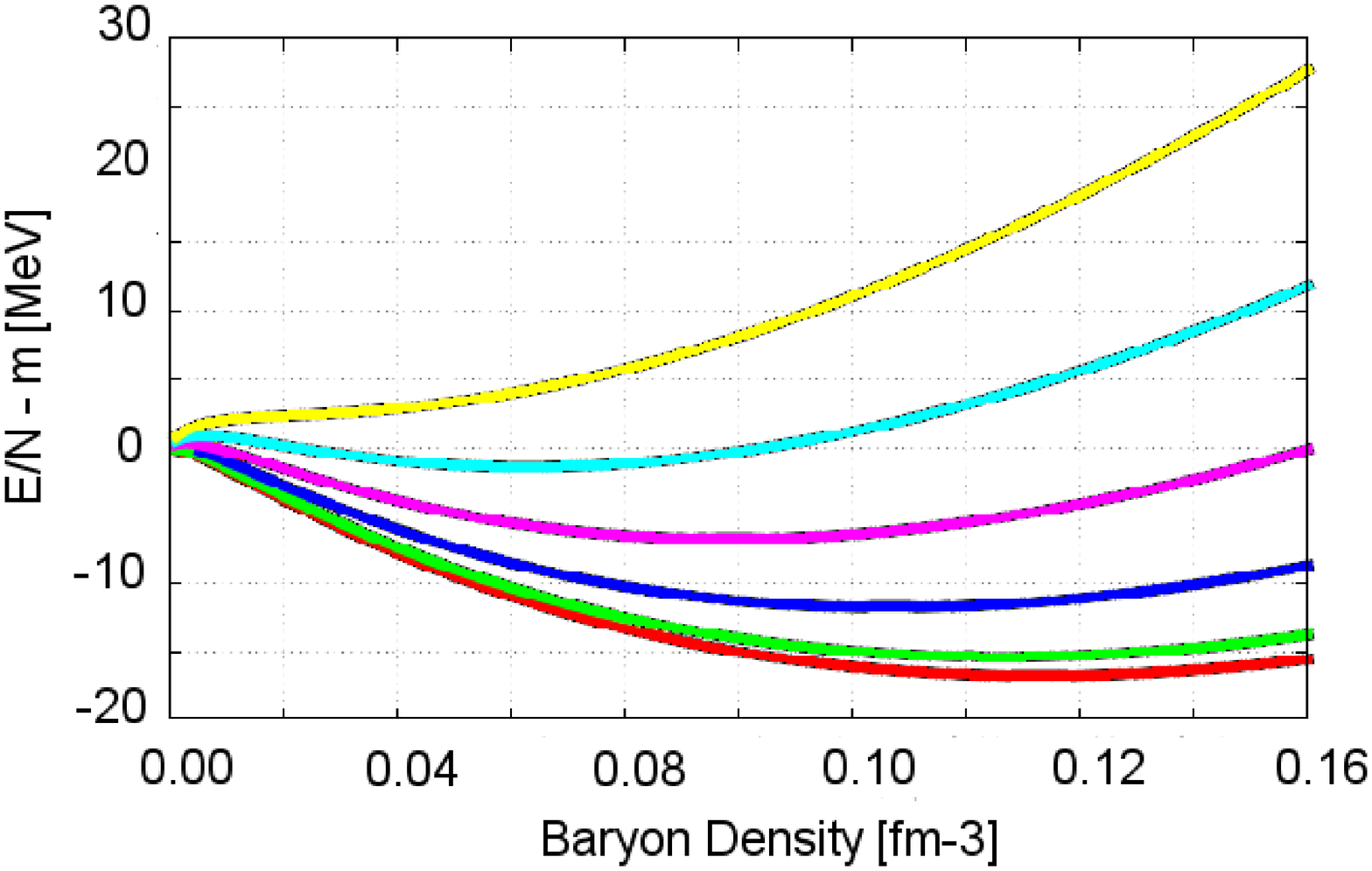}
  }
 \hspace{6pt}
 \subfigure[]{
 \includegraphics[width=0.4\textwidth]{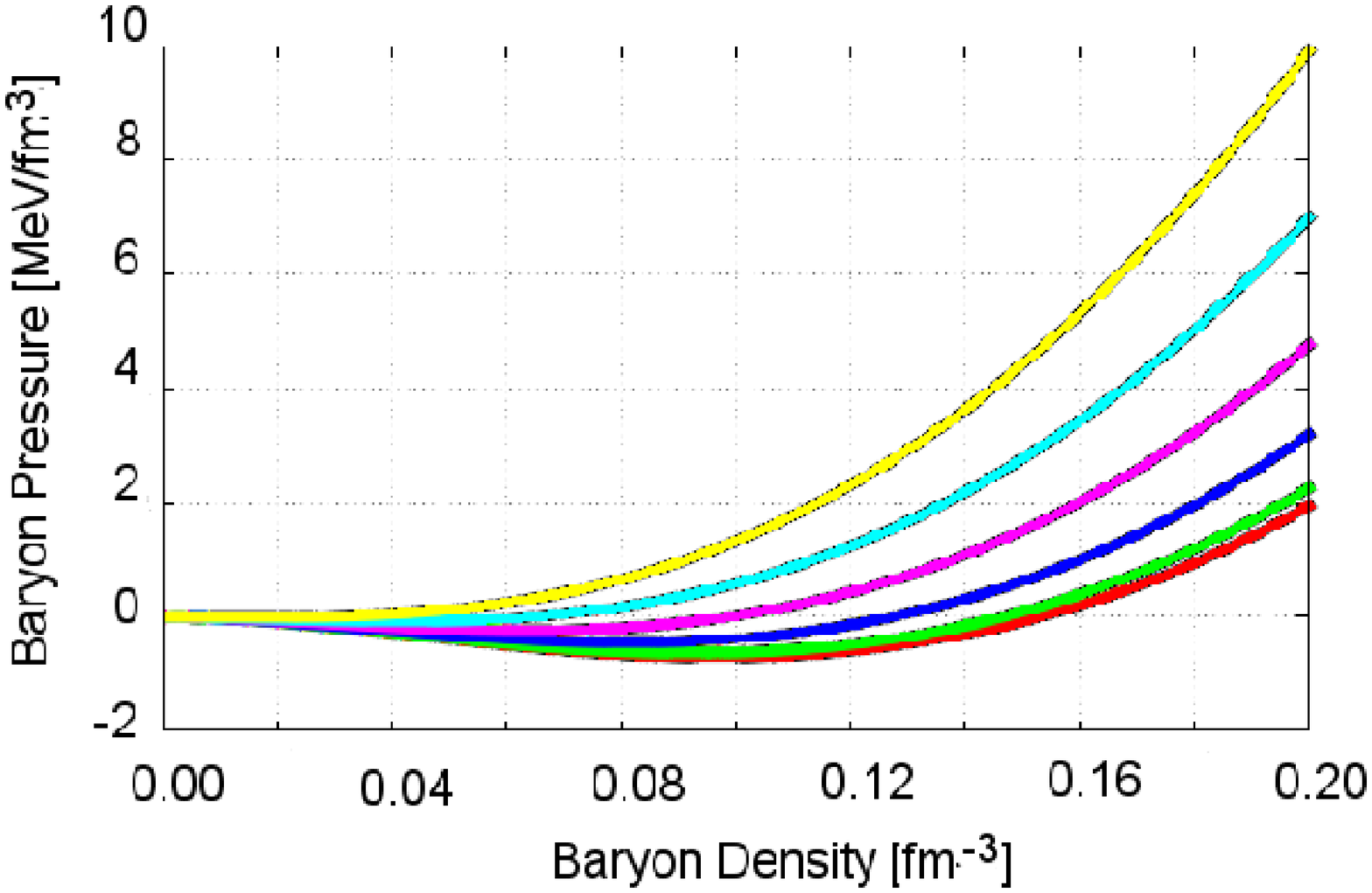}
  }
 \caption{
(a)  Binding energy and (b) pressure without the contribution of electrons.
Red line corresponds to the case of proton mixing ratio ${Y}_p=0.5$ and green, blue,
purple, sky blue and yellow correspond to 0.4, 0.3, 0.2, 0.1 and 0, respectively.
\label{fig_uniform_EOS_ne}}
 \end{center}
\label{fig_uniform_EOS}
\end{figure}

We use the same parameter set of Ref.\ \cite{relamaruyama}, 
which reproduces the saturation property of nuclear matter 
(Fig.\ \ref{fig_uniform_EOS_ne}), 
and the binding energies and the proton mixing ratios of finite nuclei.
We list the parameter set in Table~\ref{parameter set}.

\begin{table}[h]
\begin{center}
\begin{tabular}{cccccccc}
\hline
$g_{{\sigma}N}$ & $g_{{\omega}N}$ & $g_{{\rho}N}$ &
b & c & $m_{\sigma}\left({\rm MeV}\right)$ &
$m_{\omega}\left({\rm MeV}\right)$ &
$m_{\rho}\left({\rm MeV}\right)$ \\
\hline
6.3935 & 8.7207 & 4.2696 & 0.008659 & $-0.002421$ & 400 & 783 & 769 \\
\hline
\end{tabular}
\caption{Parameter set \label{parameter set}}
\end{center}
\end{table}

\subsection{Numerical Calculation}

To simulate infinite matter, we employ a periodic boundary condition 
to the calculation cell with a cubic shape.
Desirable cell size is large as possible. 
We divide the cell into grid points.
The density distributions and the meson-field profiles 
are represented by their local values on the grid points. 
Giving an average densities of protons, neutrons and electrons, but 
density distributions are randomly provided.
The suitable density distributions and fields are searched for.

The meson fields and the Coulomb potential are obtained by
solving the equations of motion (\ref{sigma_eq1})-(\ref{Coulomb_eq}).
To obtain the density distributions of nucleons and electrons 
we introduce local chemical potentials.
The equilibrium state is determined 
so that the chemical potentials are independent of the position. 
An exception is the region with a particle density equals to zero, 
where the chemical potential of that particle becomes higher.

\begin{eqnarray}
 \mu_p &=& \sqrt{\left(3{\pi}^2{\rho}_p\left(\mbox{\boldmath$r$}\right)\right)^{2/3}+\left(m_N^*\left(\mbox{\boldmath$r$}\right)\right)^2}
          +g_{{\omega}N}{\omega}_0\left(\mbox{\boldmath$r$}\right)
          +g_{{\rho}N}{R}_0\left(\mbox{\boldmath$r$}\right)
          -V\left(\mbox{\boldmath$r$}\right) \\
 \mu_n &=& \sqrt{\left(3{\pi}^2{\rho}_n\left(\mbox{\boldmath$r$}\right)\right)^{2/3}+\left(m_N^*\left(\mbox{\boldmath$r$}\right)\right)^2}
          +g_{{\omega}N}{\omega}_0\left(\mbox{\boldmath$r$}\right)
          -g_{{\rho}N}{R}_0\left(\mbox{\boldmath$r$}\right) \\
 \mu_e &=& \left(3{\pi}^2{\rho}_e\left(\mbox{\boldmath$r$}\right)\right)^{1/3} + V\left(\mbox{\boldmath$r$}\right)
\label{chemical_potential_e}
\end{eqnarray}

Starting from given density distributions, we repeat the following procedures to 
attain uniformity of chemical potentials.

\begin{center}
\fbox{
\begin{tabular}{l}
1. Calculate the chemical potentials on all of the grid points.\\
2. Compare chemical potentials on the neighbor six grid points.\\
3. If the chemical potential of the point under consideration \\
is larger than that of another, give some part of the density to \\
 the other grid point.
\end{tabular}
}
\label{method_baryon_tab}
\end{center}

\subsection{Coulomb Energy}
In the Coulomb and electron energies (\ref{Total_Energy_Coulomb}),
we consider the energy only within the calculating cell without
interaction with the neighbors. 
Therefore we do not include the Coulomb energy of the higher order,
although we solve the Poisson equation completely. 
The dipole interaction which occupies the biggest contribution 
among the higher order interactions, comes up to 5$\%$ of the Coulomb energy.
So we subtract the dipole moment of the electric field 
by carrying out a translation of the coordinate 
so that the dipole moment of charge density in the cell diminishes.

\section{Results}

We calculate three-dimensional structures of low-density nuclear matter 
at zero temperature and obtain the energy or pressure as function of density, 
i.e.\ the equation of state (EOS).
In this report, let us first demonstrate the cases with the fixed proton
mixing ratio, $Y_p\equiv N_p/N_B$, 
and then with beta-equilibrium.
Particularly, we set the proton mixing ratio to ${Y}_p=$0.5, 0.3 and 0.1.
Because these cases should be the one of typical nuclear matter, 
relevant for the supernova core, and for protoneutron stars. 

In some of the previous works which have used the WS cell approximation,
nuclear matter might be enforced to have some typical pasta structures.
In reality, however, there might appear some intermediate shapes in the density regions where
structure changes. 
With the WS cell approximation, the cell size which gives the minimum energy density
has been calculated \cite{relamaruyama}.
To save the computational effort, we make use of that size of the WS cell 
for the size of our cell with a periodic boundary condition.

\begin{figure}[H]
 \begin{center}
  \subfigure[droplet]{
  \includegraphics[width=2cm,height=2cm]{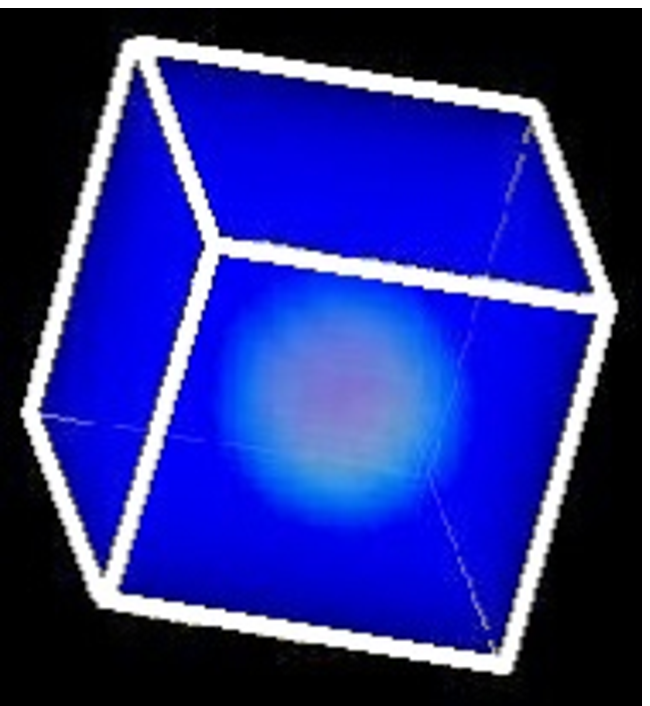}
  }
  \subfigure[rod]{
  \includegraphics[width=2cm,height=2cm]{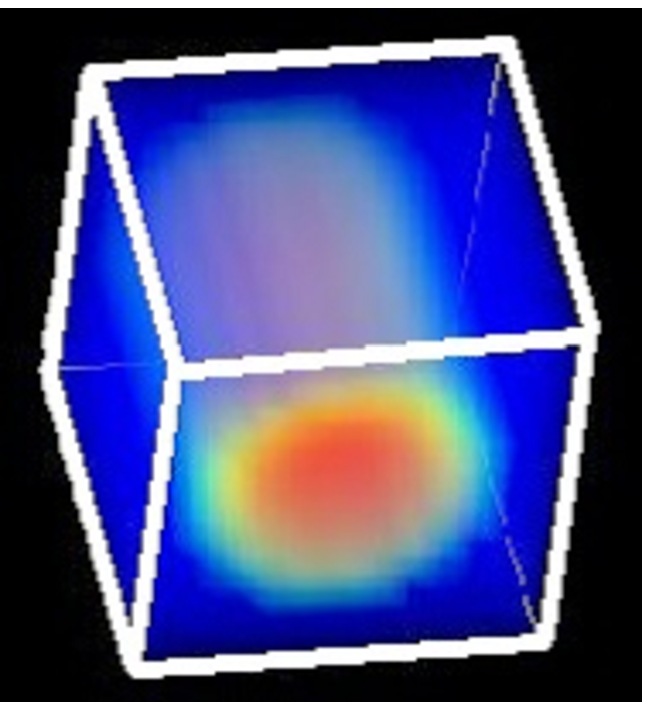}
   }
  \subfigure[slab]{
  \includegraphics[width=2cm,height=2cm]{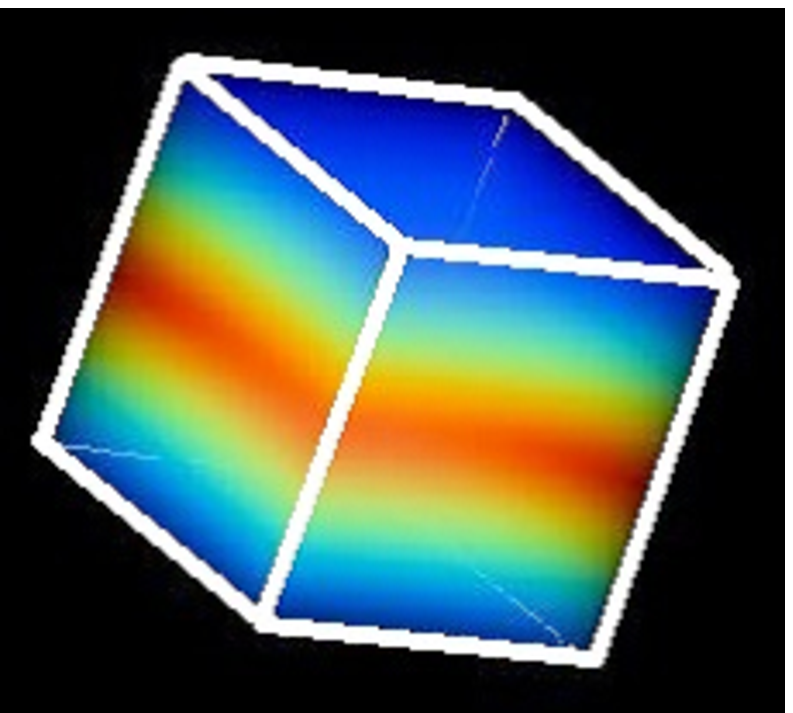}
   }
  \subfigure[tube]{
  \includegraphics[width=2cm,height=2cm]{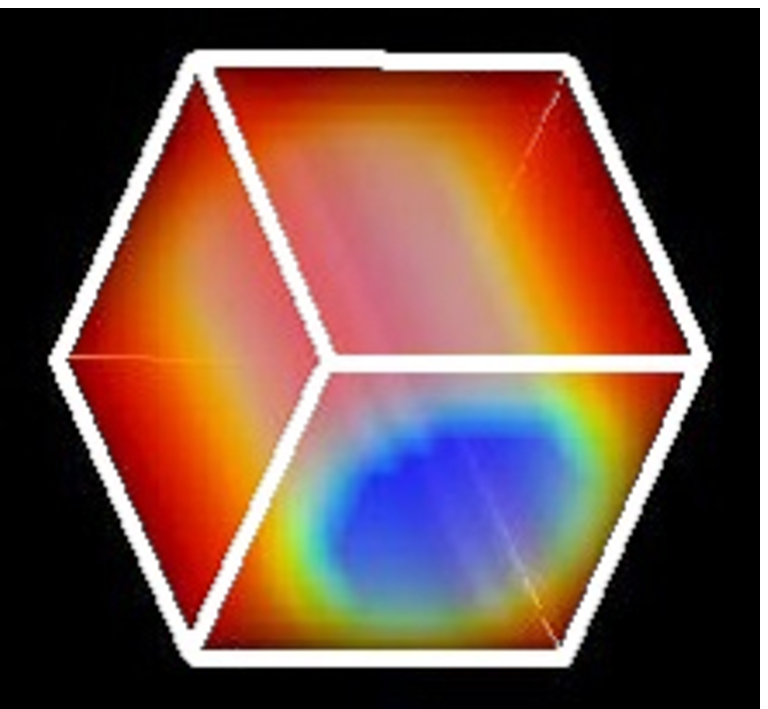}
   }
  \subfigure[bubble]{
  \includegraphics[width=2cm,height=2cm]{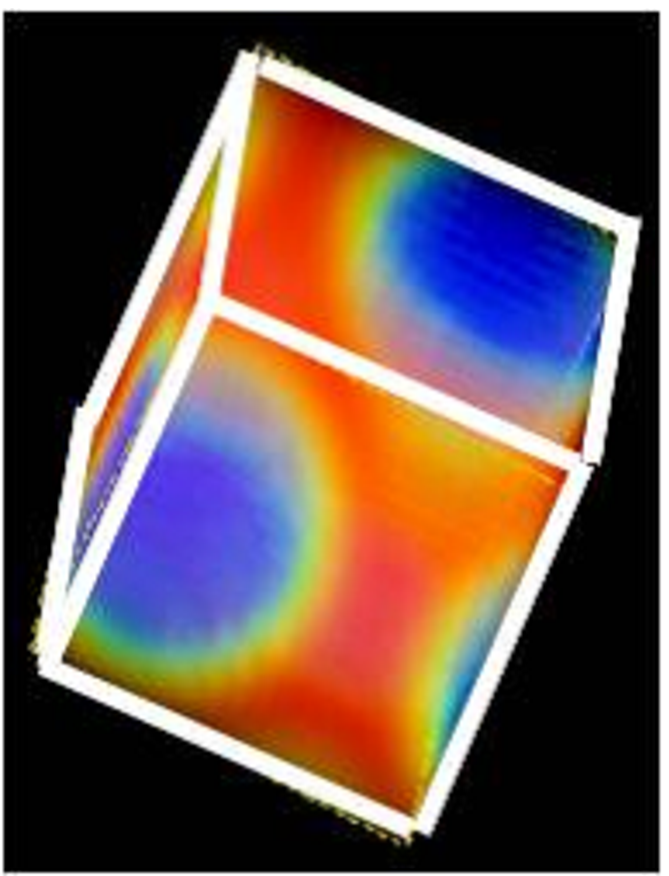}
   }
  \subfigure[uniform]{
  \includegraphics[width=2cm,height=2cm]{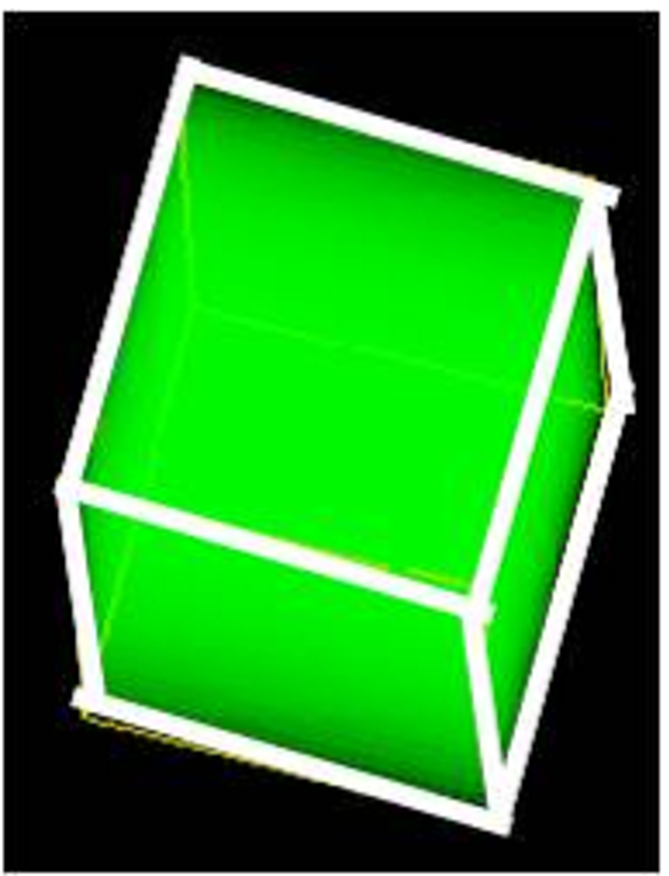}
  }
 \end{center}
  \caption{Proton density distribution for symmetric nuclear matter (${Y}_{p}=0.5$)
  \protect\newline
  Red color corresponds to the highest density 0.08 fm$^{-3}$,
  and blue corresponds to 0 fm$^{-3}$. \label{fig Yp=05 proton}}
\end{figure}

\begin{figure}[H]
 \begin{center}
  \subfigure[Binding energy]{
  \includegraphics[width=0.3\textwidth]{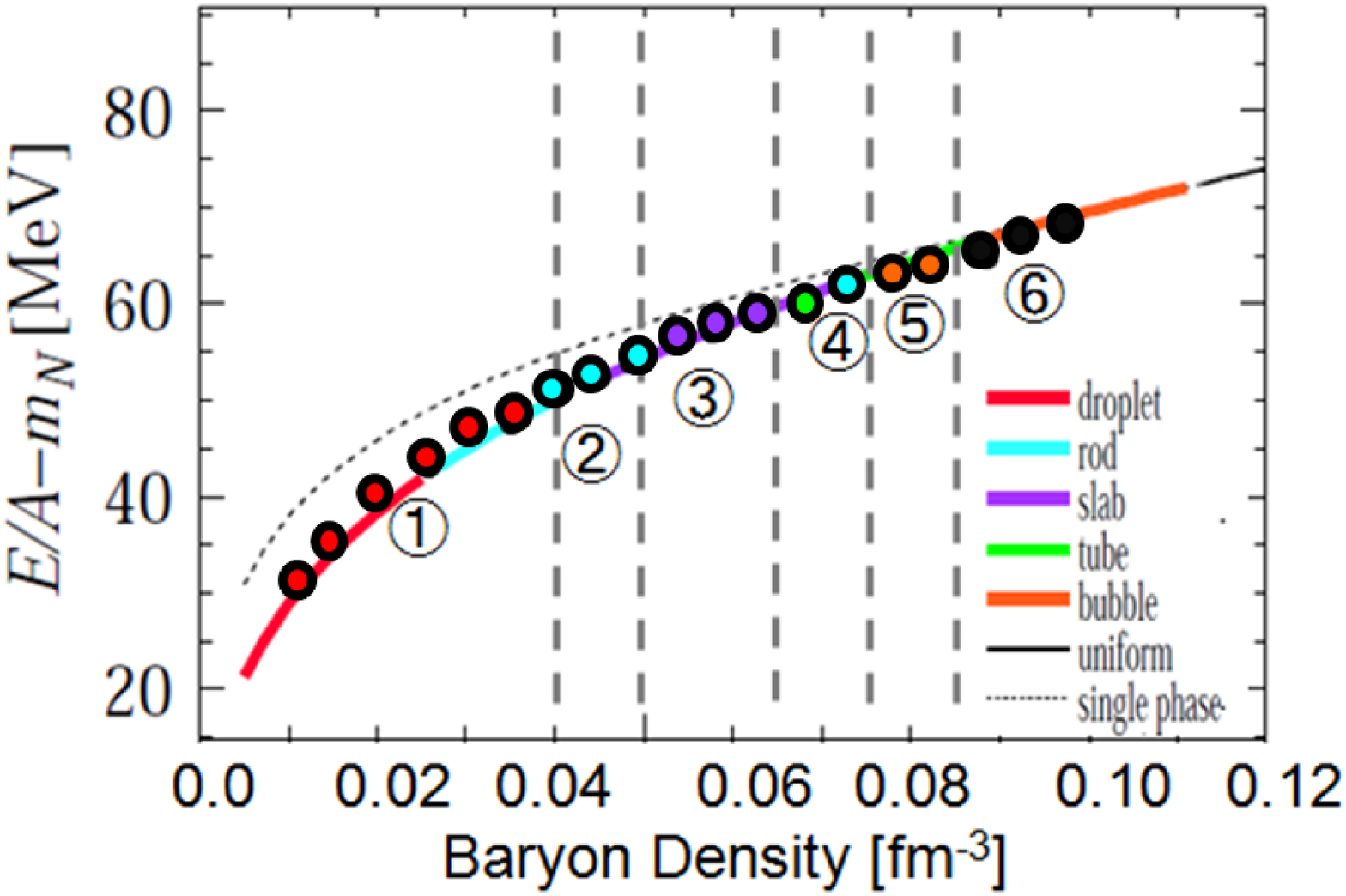}}
  \label{fig EOS Yp=05}
  \subfigure[Total pressure]{
  \includegraphics[width=0.3\textwidth]{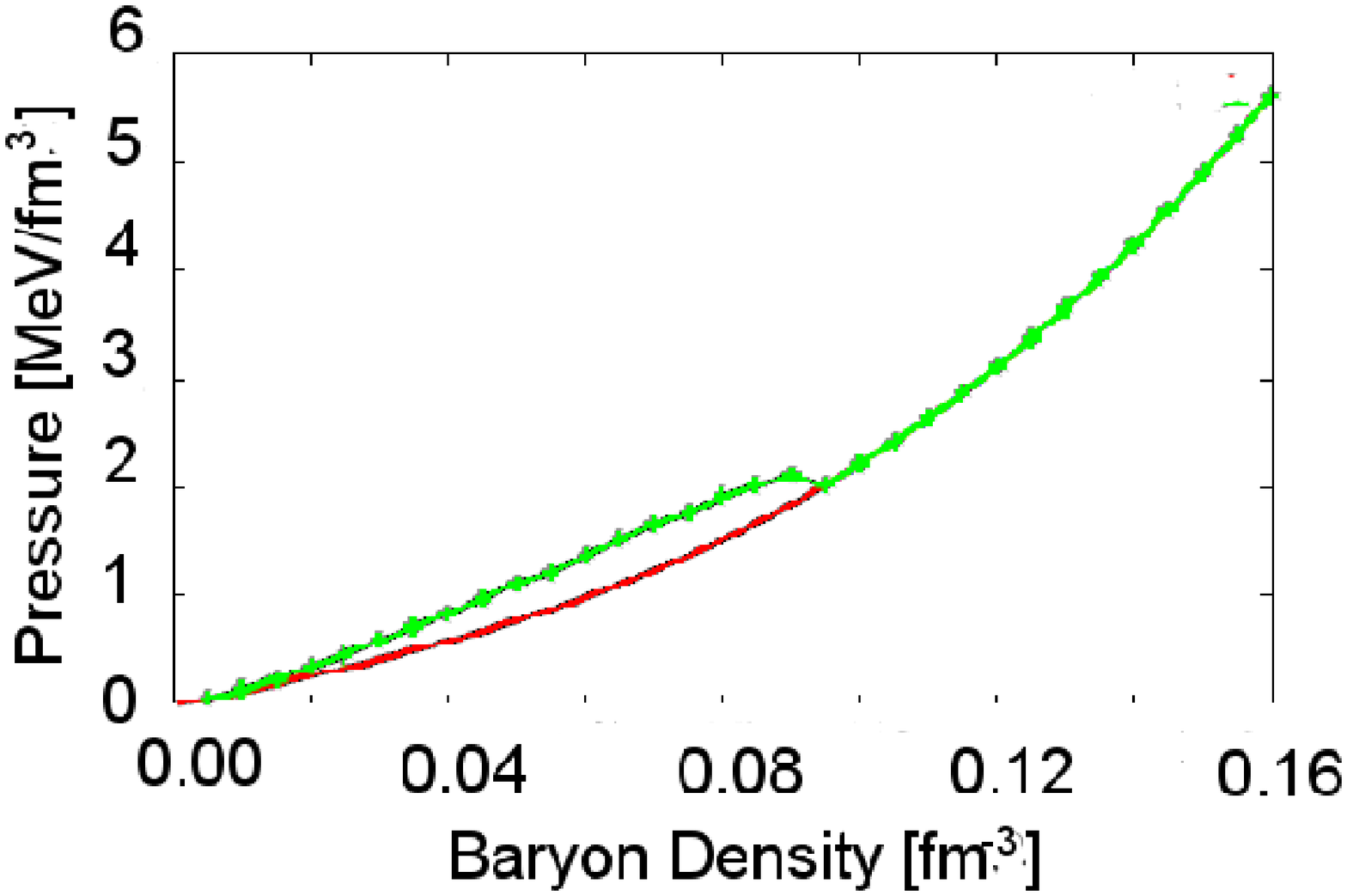}}
  \label{Pressure Yp=05}
  \subfigure[Baryon pressure]{
  \includegraphics[width=0.3\textwidth]{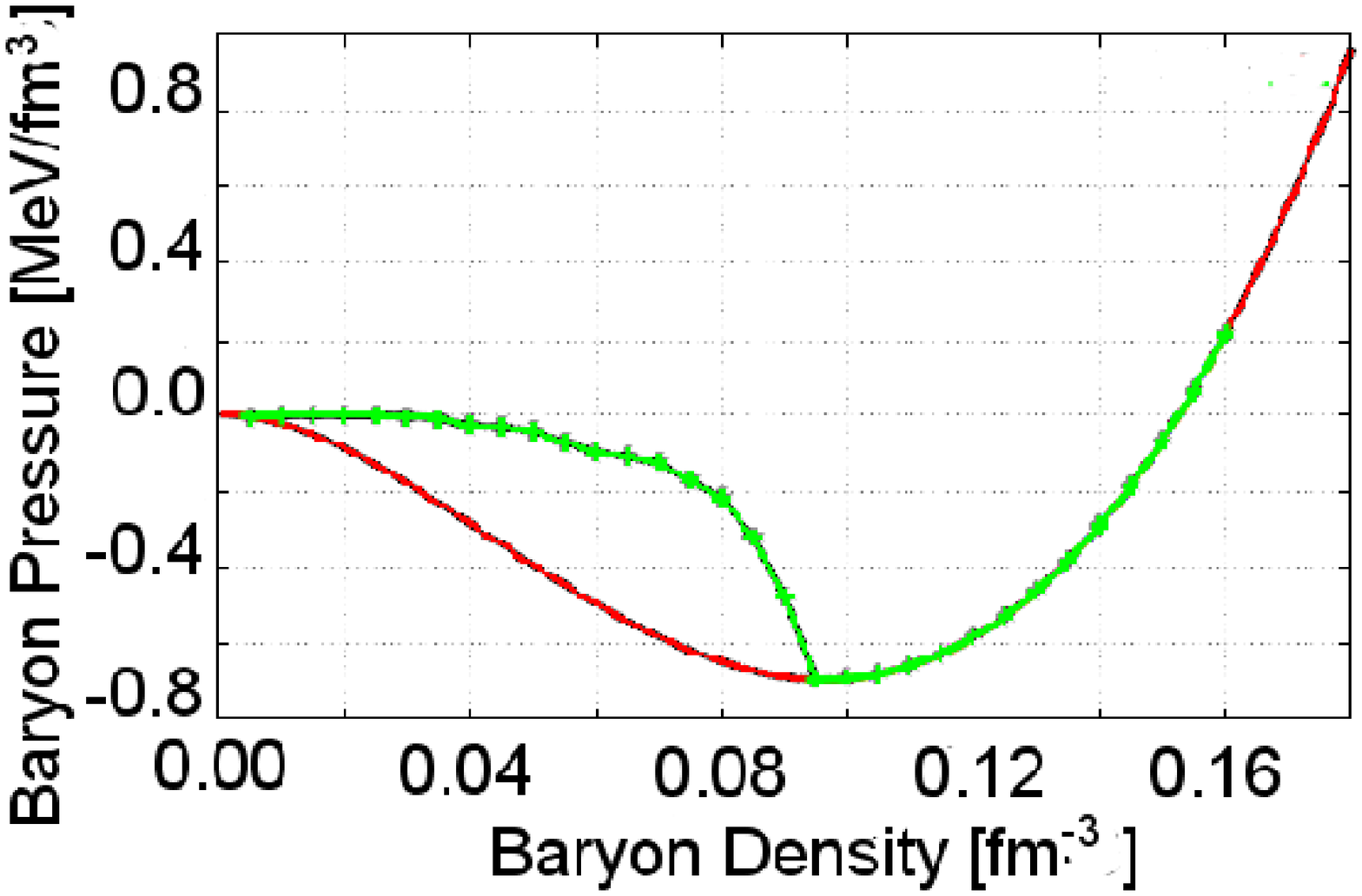}}
  \label{Baryon Pressure Yp=05}
  \end{center}
 \caption{EOS of matter with $Y_p=0.5$.
     (a) Energy per nucleon obtained by the present calculation (colored circles)
compared with the results of Ref.\ \cite{relamaruyama}.
      (b) Total pressure and (c) baryon partial pressure.
Black dotted lines in (a), (b) and (c) show the cases of uniform matter. 
\label{EOS Yp=05}}
\end{figure}

First, we show the result for symmetric nuclear matter 
with a proton mixing ratio ${Y}_p=0.5$ in Fig.\ \ref{fig Yp=05 proton}.
Panels (a), (b), (c), (d), (e), and (f) correspond to droplet, rod, slab, tube,
bubble and uniform, respectively. 
In our calculation, all of the typical
pasta structures are seen.

The binding energy, the total pressure and the
 baryon partial pressure are presented in Fig.\ \ref{EOS Yp=05}.
The line with colors, dashed line and dots correspond to 
the results with the WS cell approximation, 
the case of uniform matter, and our results 
by the three-dimensional calculation, respectively.
The density region with numbers $1, 2, \cdots, 6$ 
separated with vertical dashed lines
indicate that the structures 
in Fig.\ \ref{fig Yp=05 proton} (a), (b), $\cdots$, (f) appear.
Note that the density range for each pasta structure is slightly different
from the previous study with the WS cell approximation.

\begin{figure}[H]
 \begin{center}
  \subfigure[droplet]{
   \includegraphics[width=2cm,height=2cm]{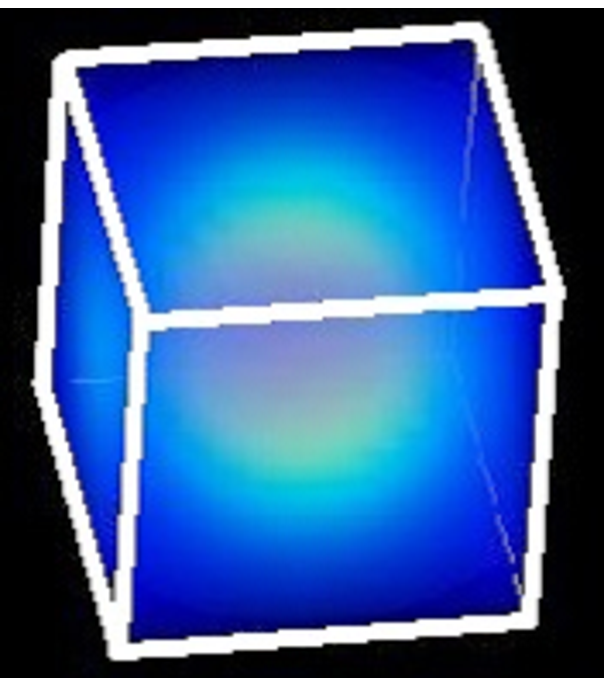}
   }
  \subfigure[rod]{
   \includegraphics[width=2cm,height=2cm]{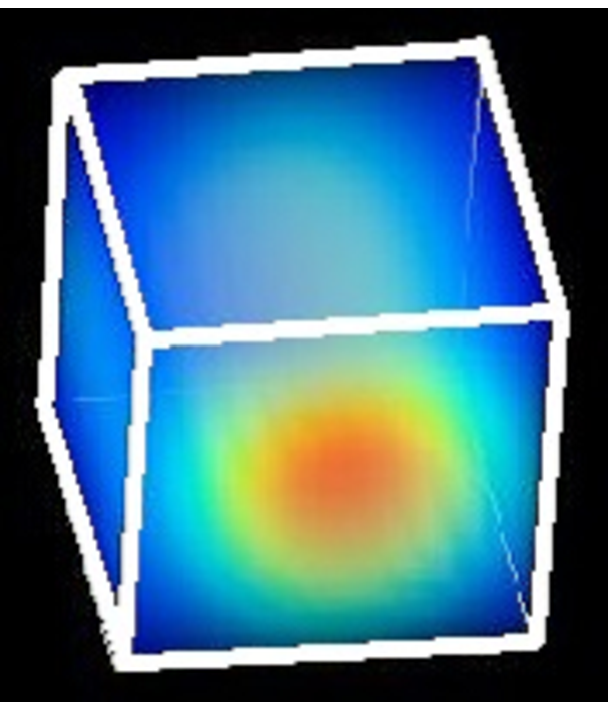}
   }
  \subfigure[slab]{
   \includegraphics[width=2cm,height=2cm]{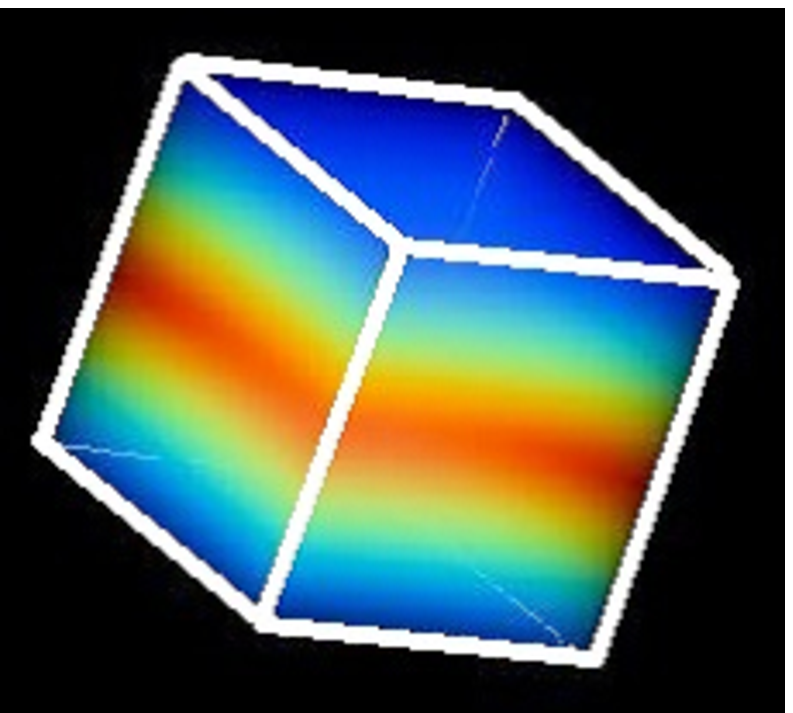}
   }
  \subfigure[tube]{
   \includegraphics[width=2cm,height=2cm]{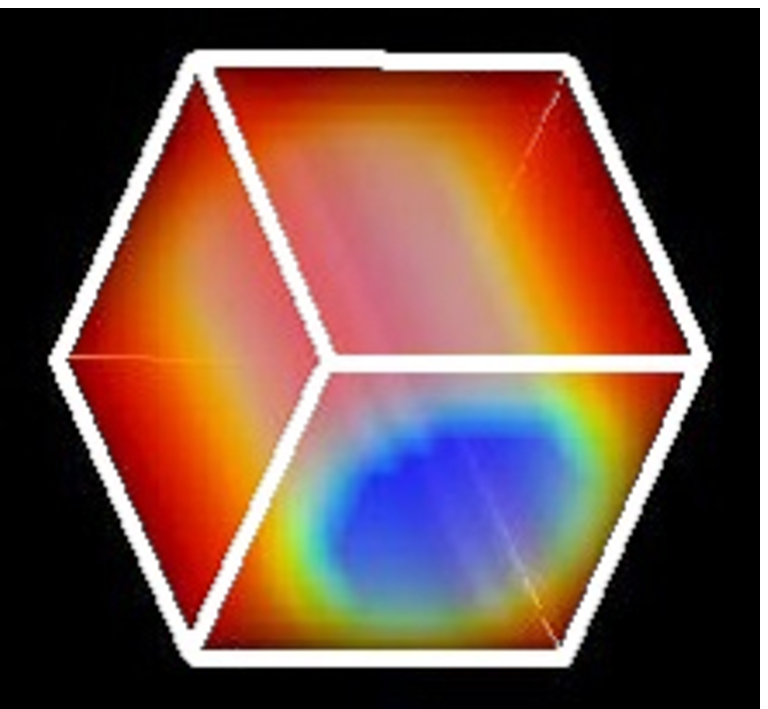}
   }
  \subfigure[uniform]{
   \includegraphics[width=2cm,height=2cm]{fig13.eps}
   }
 \end{center}
 \caption{Same as Fig.\ \ref{fig Yp=05 proton} for ${Y}_{p}=0.3$.
 But the density range from blue to red is 0 to 0.075 fm$^{-3}$.}
\end{figure}

\begin{figure}[H]
 \begin{center}
  \subfigure[Binding energy] {
  \includegraphics[width=0.3\textwidth]{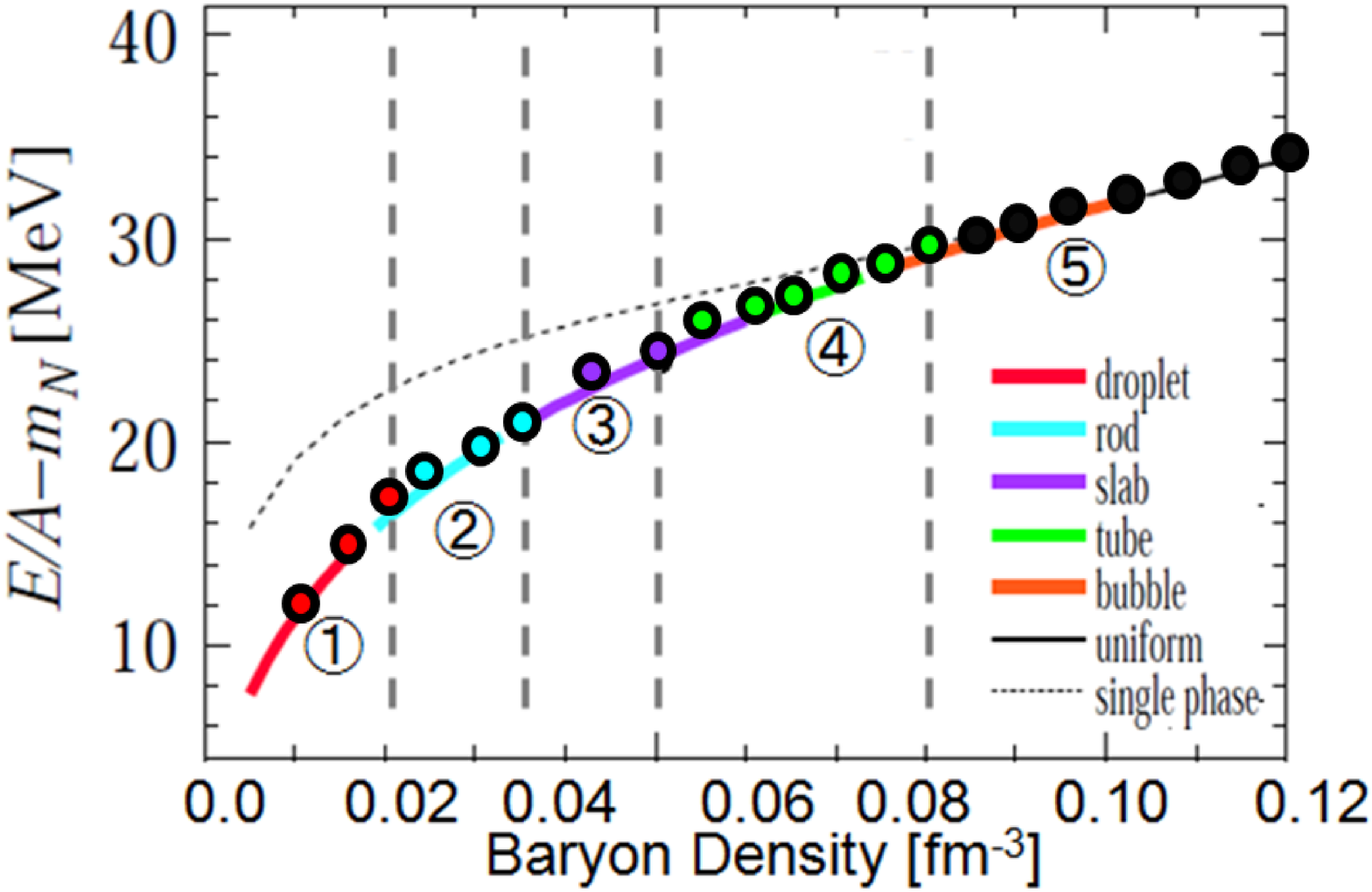}}
  \label{fig EOS Yp=03}
  \subfigure[Total pressure] {
  \includegraphics[width=0.3\textwidth]{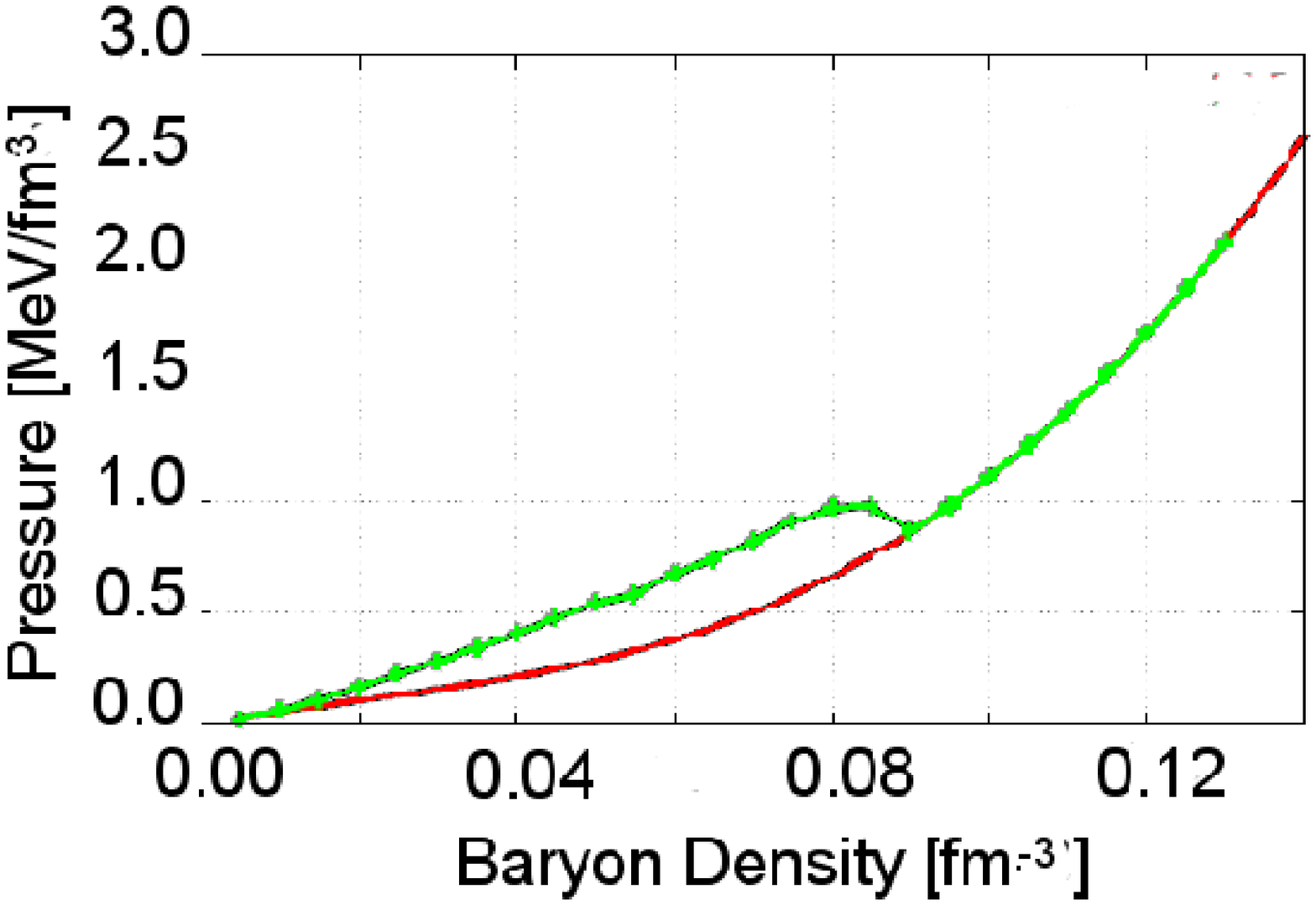}}
  \label{Pressure Yp=03}
  \subfigure[Baryon pressure]{
  \includegraphics[width=0.3\textwidth]{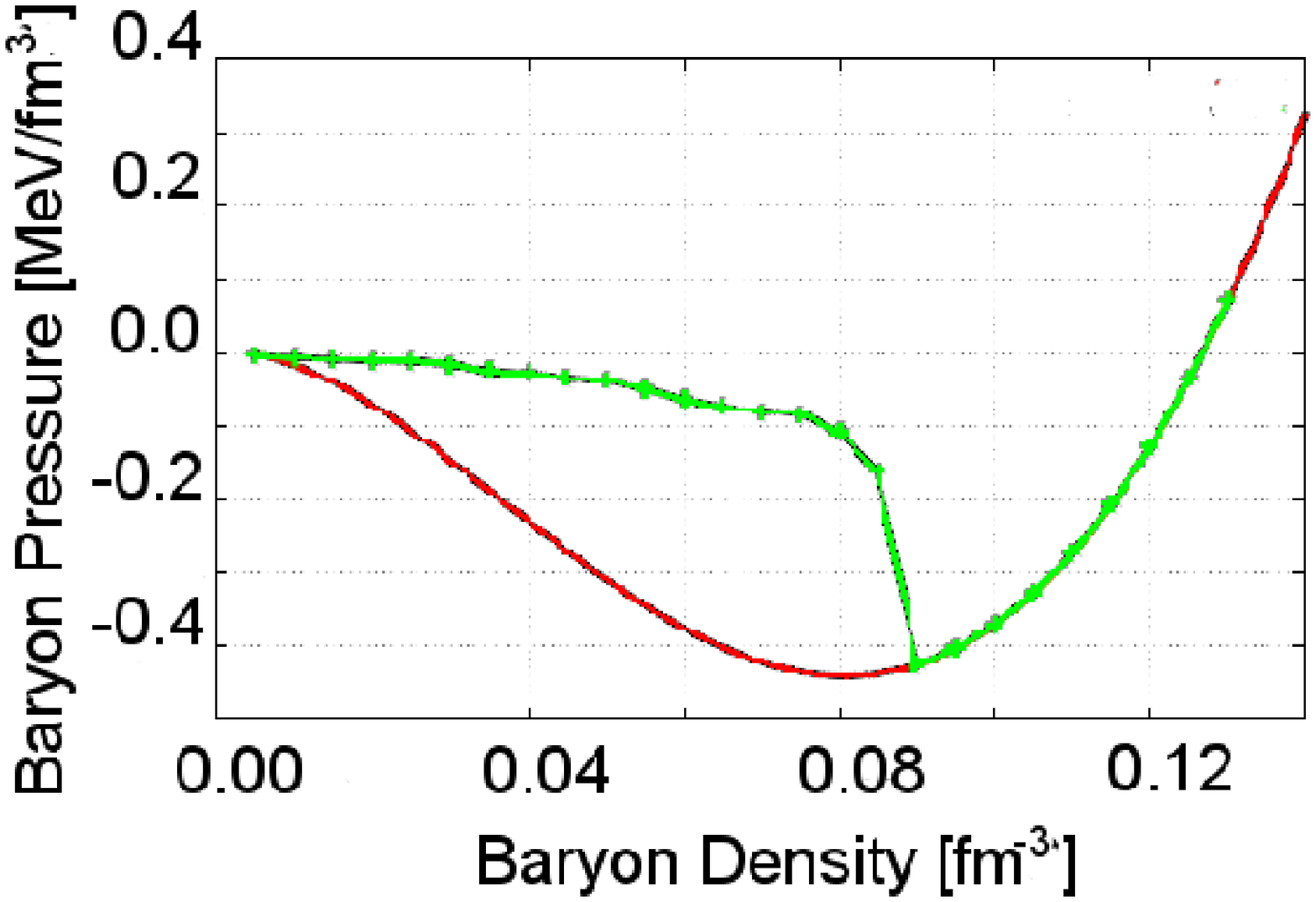}}
  \label{Baryon Pressure Yp=03}
 \end{center}
 \caption{Same as Fig.\ \ref{EOS Yp=05} for $Y_p=0.3$. \label{EOS Yp=03}}
\end{figure}

Next, we show the result for proton mixing ratio ${Y}_p=0.3$.
in Fig.\ \ref{EOS Yp=03}.
The cell size is set to be the same as in Ref.\ \cite{relamaruyama}. 
The difference between two results is that
the bubble structure does not appear in the present calculation.

Comparing the cases of $Y_p=0.5$ and $0.3$, 
the upper limit of the density where non-uniform structures appear is
different.
In the case of symmetric nuclear matter, the density region of 
non-uniform matter roughly corresponds to the spinodal region, where $dP/d\rho_B<0$,  
while the non-uniform region is slightly wider than the spinodal region for $Y_p=0.3$.
This may be because the symmetric nuclear matter behaves congruently 
(phase transition in a single chemical component), while the liquid-gas 
mixed phase in matter with $Y_p=0.3$ is non-congruent and the values of $Y_p$ in 
two phases are generally different. 
Therefore the instability of uniform matter is determined not only by
the compressibility of uniform matter but also by the chemical composition 
of the mixed phase to be formed.

Next, we show the result of ${Y}_p=0.1$.
\begin{figure}[H]
 \begin{center}
  \subfigure[droplet]{
   \includegraphics[width=2cm,height=2cm]{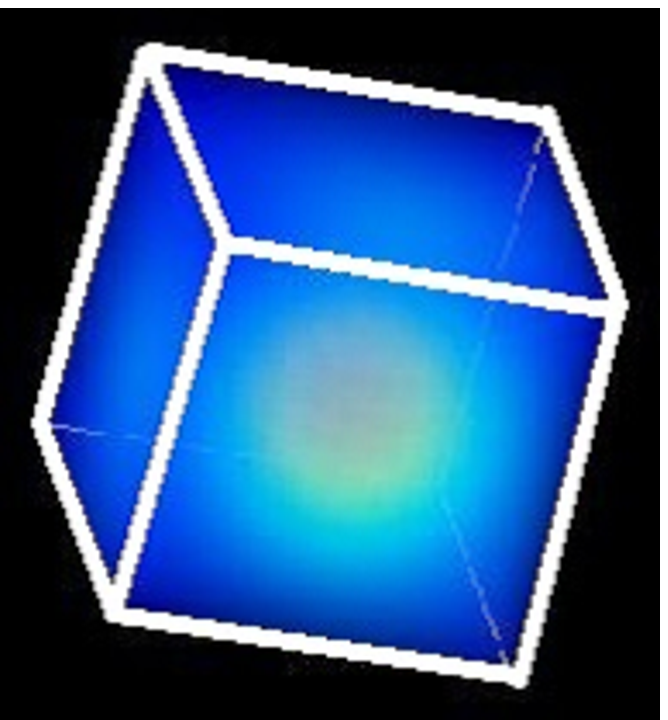}
   }
  \subfigure[rod]{
   \includegraphics[width=2cm,height=2cm]{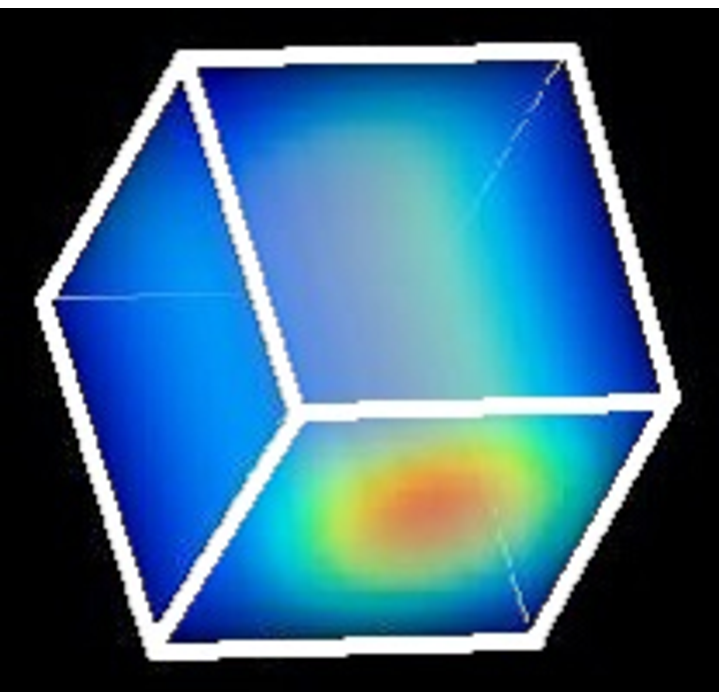}
   }
  \subfigure[slab]{
   \includegraphics[width=2cm,height=2cm]{slab_n_Yp=01.eps}
   }
 \end{center}
 \caption{Same as Fig.\ \ref{fig Yp=05 proton} for ${Y}_{p}=0.1$.
 The density range from blue to red is 0 to 0.05 fm$^{-3}$.}
\end{figure}

\begin{figure}[H]
 \begin{center}
  \subfigure[Binding energy]{
  \includegraphics[width=0.3\textwidth]{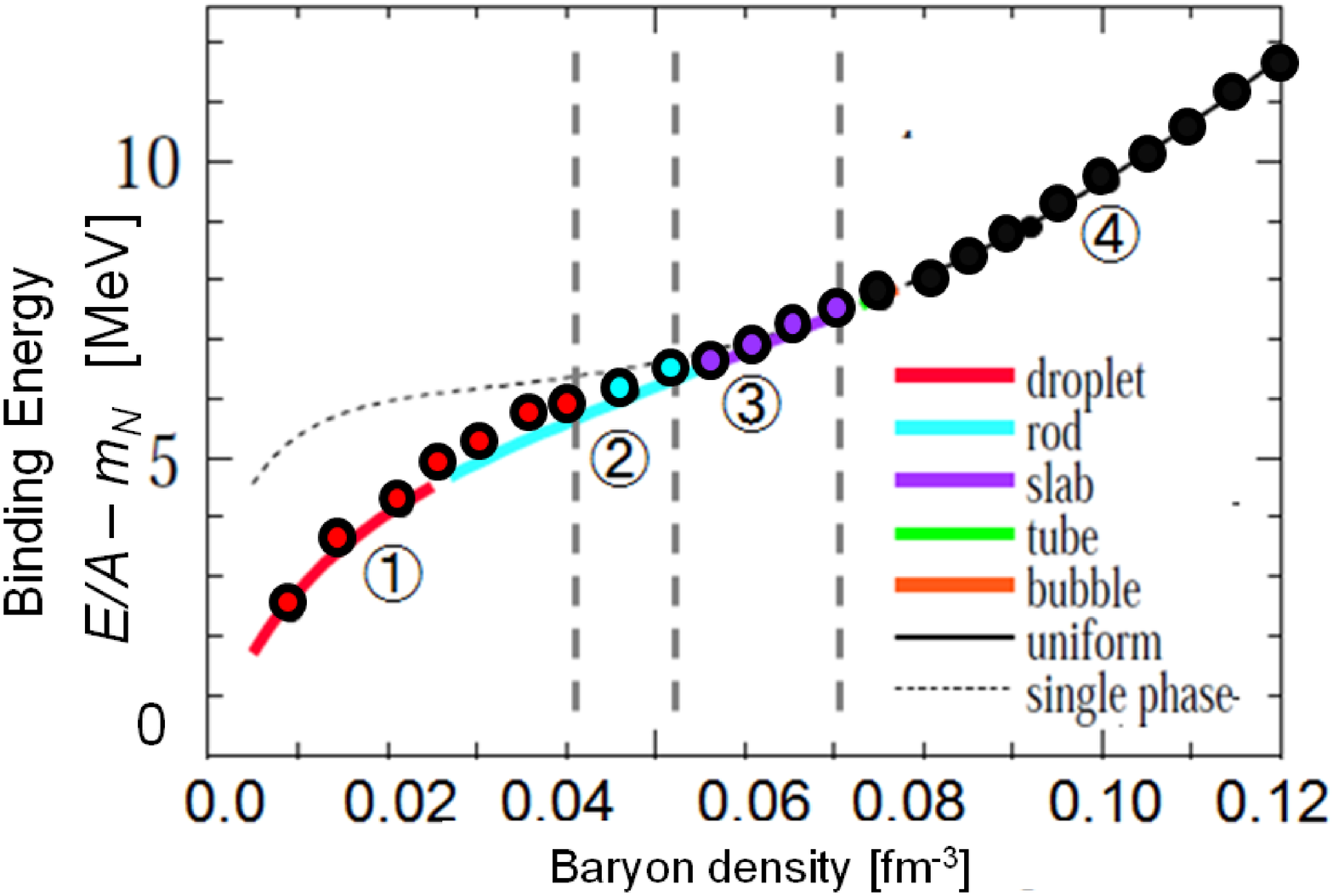}}
  \label{fig EOS Yp=01}
  \subfigure[Total pressure]{
  \includegraphics[width=0.3\textwidth]{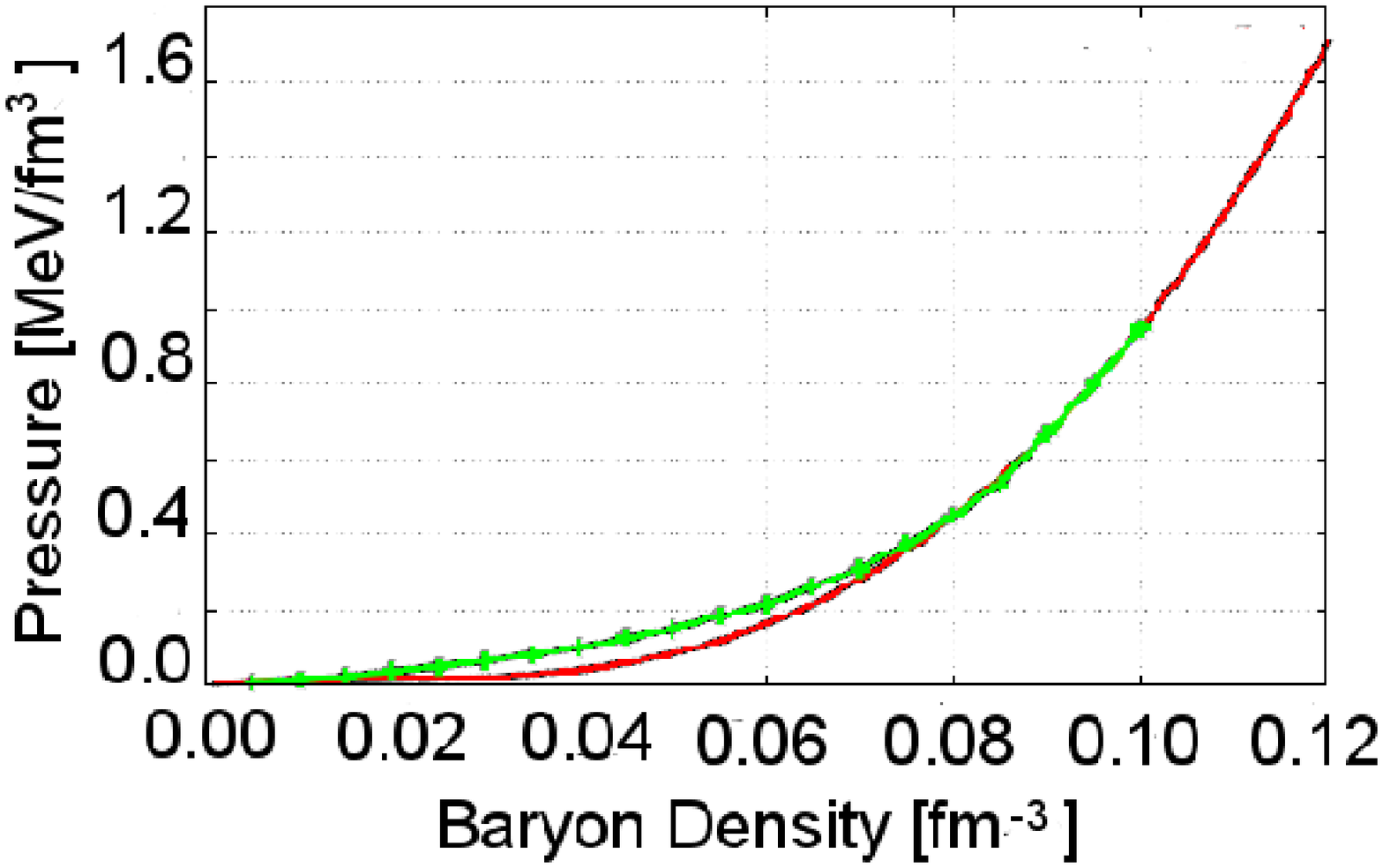}}
  \label{Pressure Yp=01}
  \subfigure[Baryon pressure]{
  \includegraphics[width=0.3\textwidth]{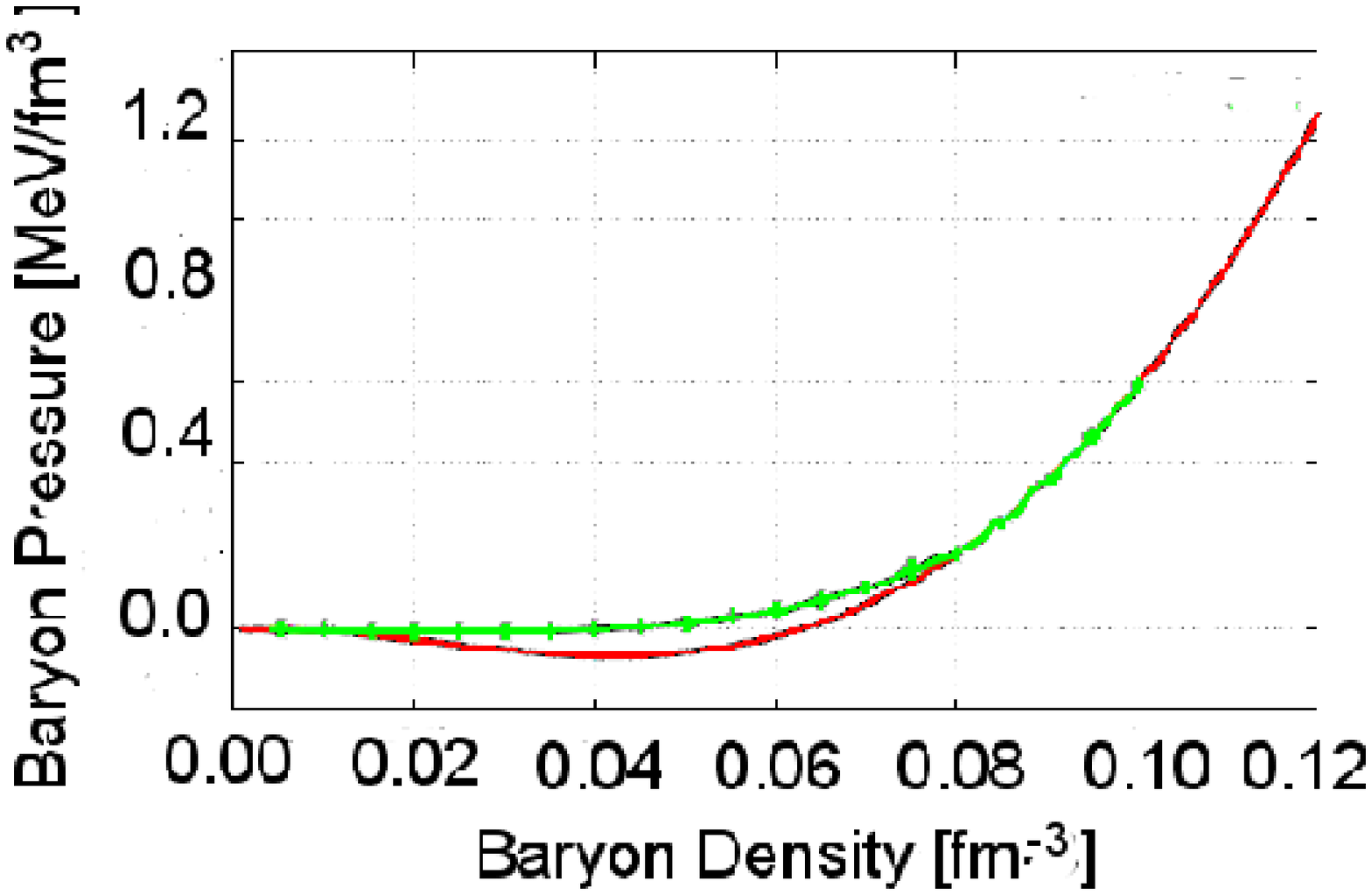}}
 \label{Baryon Pressure Yp=01}
 \end{center}
 \caption{Same as Figs.\ \ref{EOS Yp=05} and \ref{EOS Yp=03}  
for $Y_p=0.1$.\label{EOS Yp=01}}
\end{figure}

In Fig.\ \ref{EOS Yp=01}, we show the baryon density dependence of
binding energy, total pressure, and baryon partial pressure. 
We use the lines and dots in
Non-uniform structures of only droplet, rod, and slab appear.

\begin{figure}[H]
 \begin{center}
  \subfigure[$Y_p=0.5$]{
  \includegraphics[width=4cm,height=4cm]{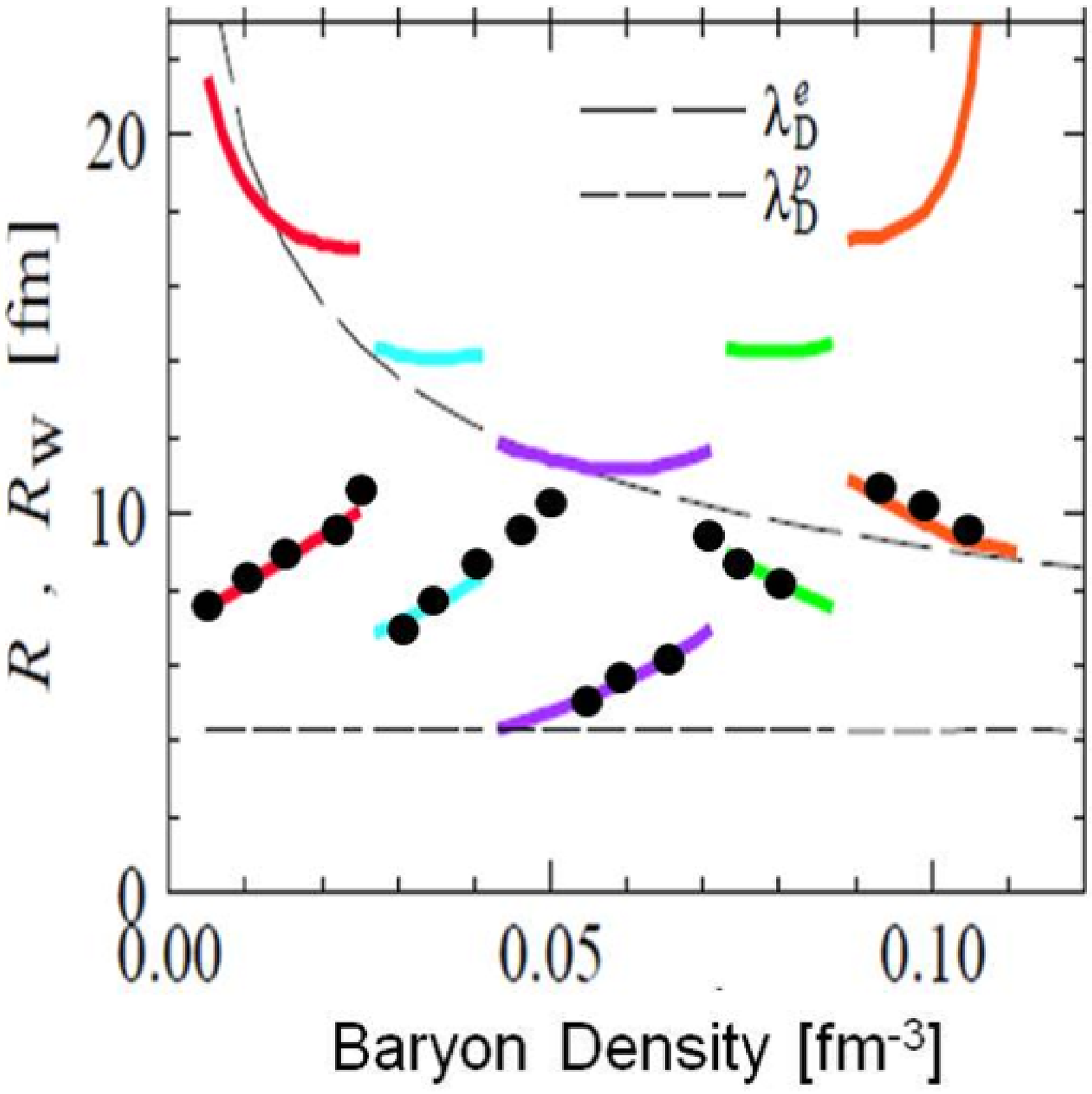}}
  \label{Yp=05_comp_ws}
  \subfigure[$Y_p=0.3$]{
  \includegraphics[width=4cm,height=4cm]{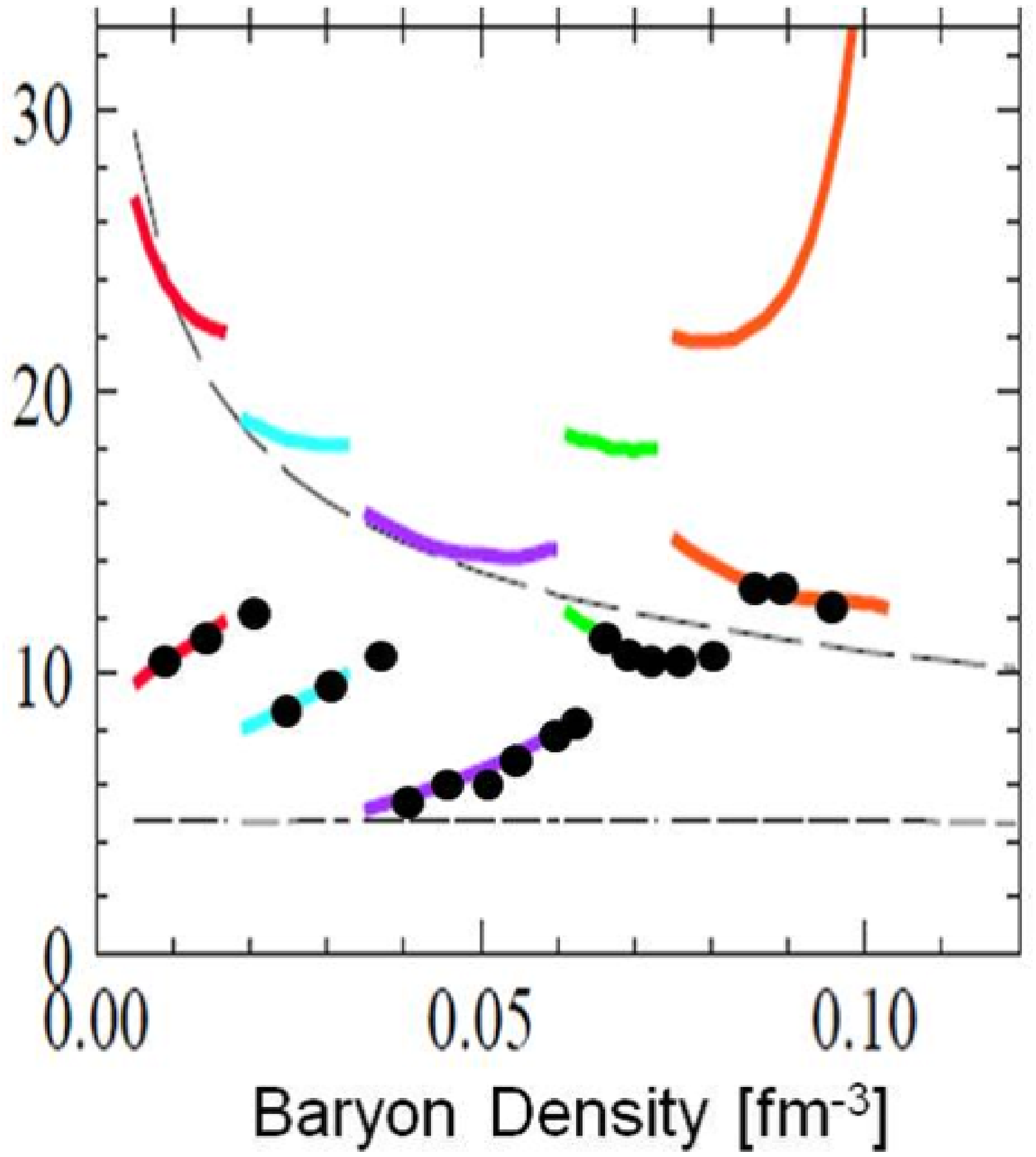}}
  \label{Yp=03_comp_ws}
  \subfigure[$Y_p=0.1$]{
  \includegraphics[width=4cm,height=4cm]{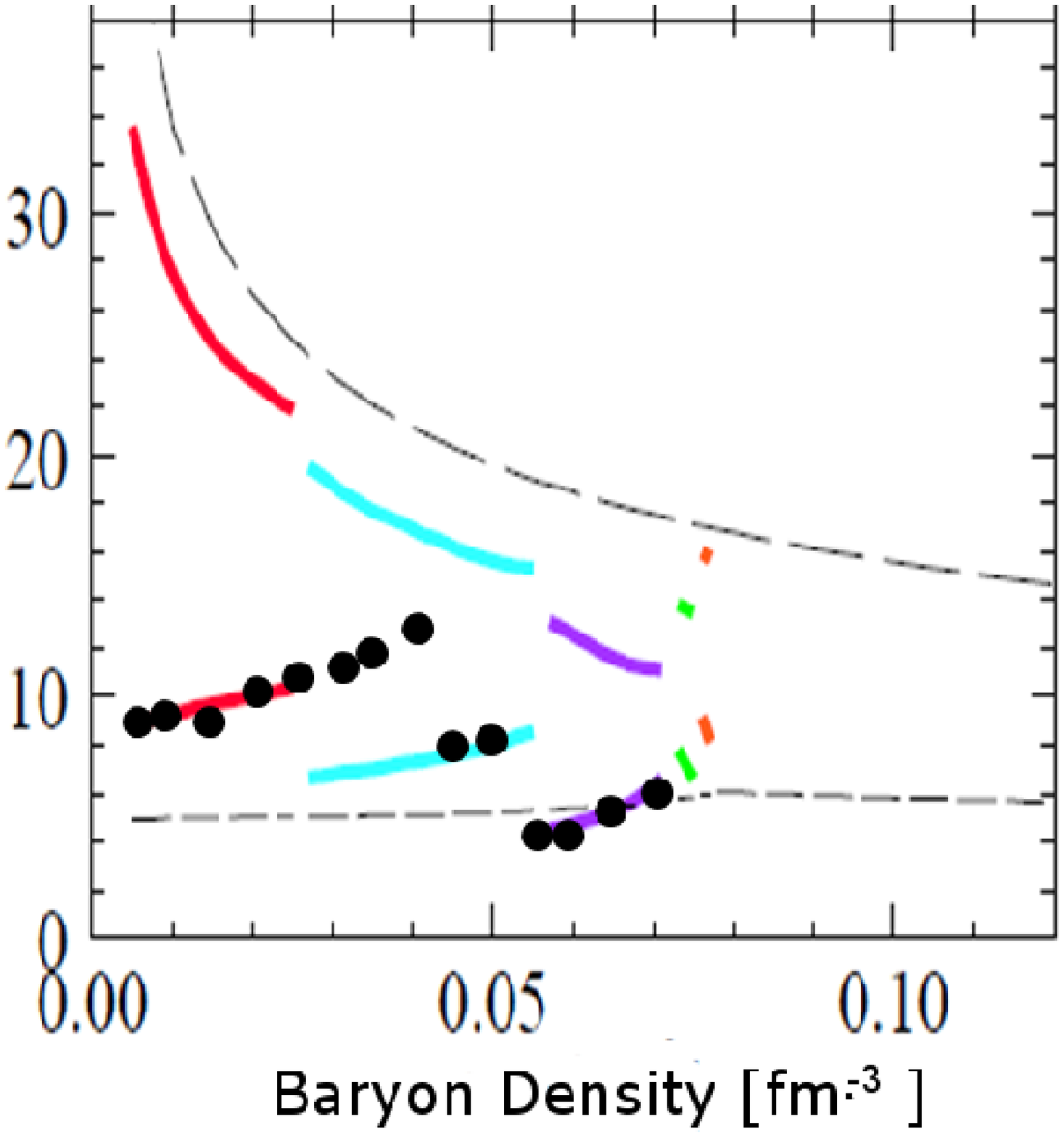}}
  \label{Yp=01_comp_ws}
 \end{center}
 \caption{Comparison of the cell and the structure size with the WS approximation 
  and three-dimensional calculation (${Y}_p$=0.3) \newline
  Line with colors : WS approximation, Dots: Three-dimensional
  calculation. Upper line with color $R_{\rm cell}$: cell size,
  Bottom line with color $R_{\rm d}$:Non-uniform structure size 
  \label{fig size with ws}}
\end{figure}
To compare quantitatively our calculation with the WS cell approximation, 
we show the cell and the structure size in Fig.\ \ref{fig size with ws}.
Upper and lower lines represent the cell and the structure
size in one cell, respectively.
Non-uniform structure size is defined by way of a density fluctuations as
\begin{equation}
 R = \left \{
      \begin{array}{l}
       R_w\frac{<{\rho}_p>^2}{<{\rho}^2_p>}, (\hbox{for droplet, rod and slab}) \\
       R_w(1-\frac{{<{\rho}_p>^2}}{{<{\rho}^2_p>}}), (\hbox{for tube and bubble})
      \end{array}
     \right.
\end{equation}

We show in Fig.\ \ref{fig size with ws} the size of
the cell and the non-uniform structure.
The density where each pasta structure appears 
is mostly in agreement with Ref.\ \cite{relamaruyama}.
However, there are large differences in the transient region of pasta structures.

In our calculation, we set, as the initial condition, the density
distributions of fermions (nucleons and electron) and mesons randomly.
So, the converged density distribution sometimes are trapped in local minimum states 
particularly at the density region where structure changes.

Let us discuss the case of beta-equilibrium at zero temperature,
 which is relevant to the realistic neutron star matter.
\begin{figure}[H]
 \begin{center}
  \subfigure[WS approx.]{
  \includegraphics[height=5cm]{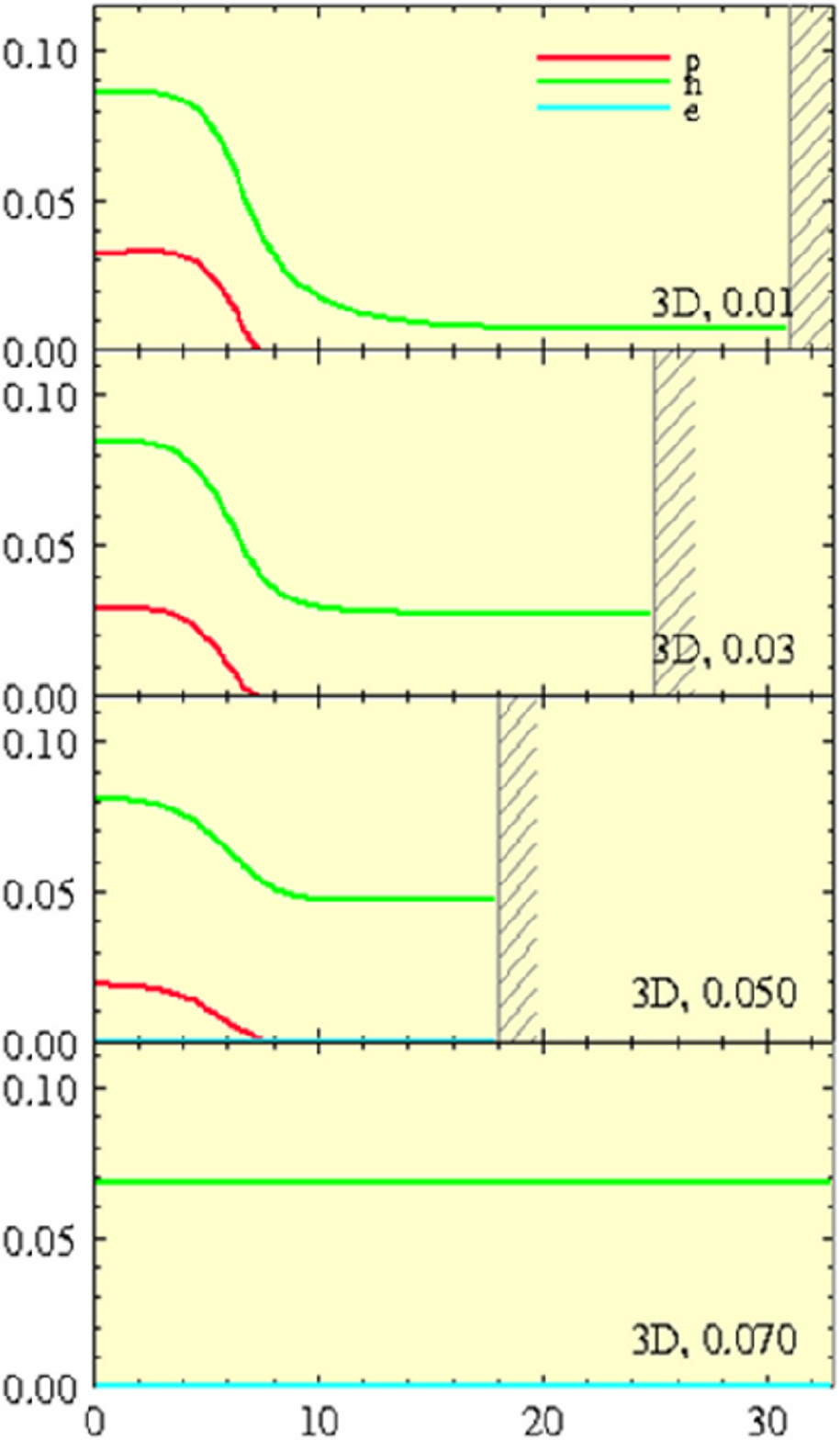}}
  \subfigure[proton]{
  \includegraphics[width=2.5cm,height=2.5cm]{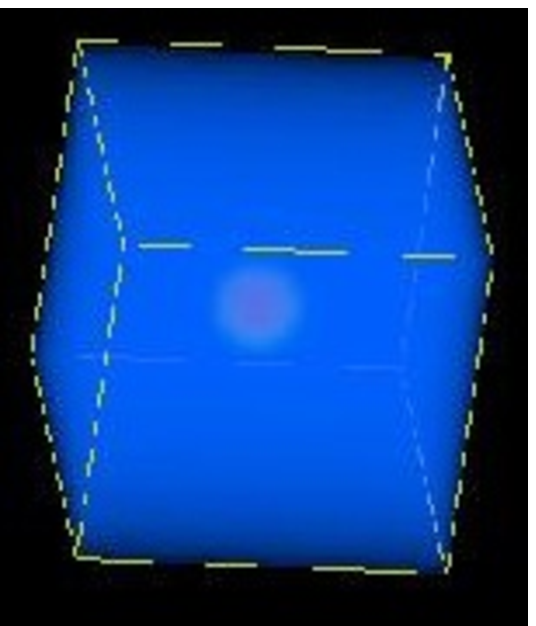}}
  \subfigure[neutron]{
  \includegraphics[width=2.5cm,height=2.5cm]{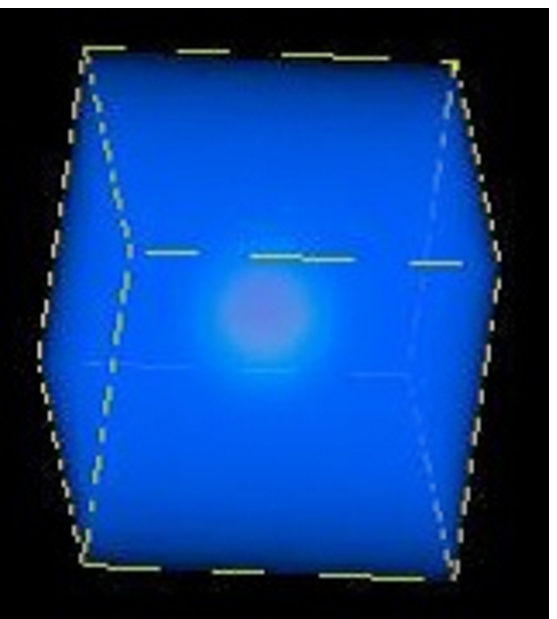}}
  \subfigure[electron]{
  \includegraphics[width=2.5cm,height=2.5cm]{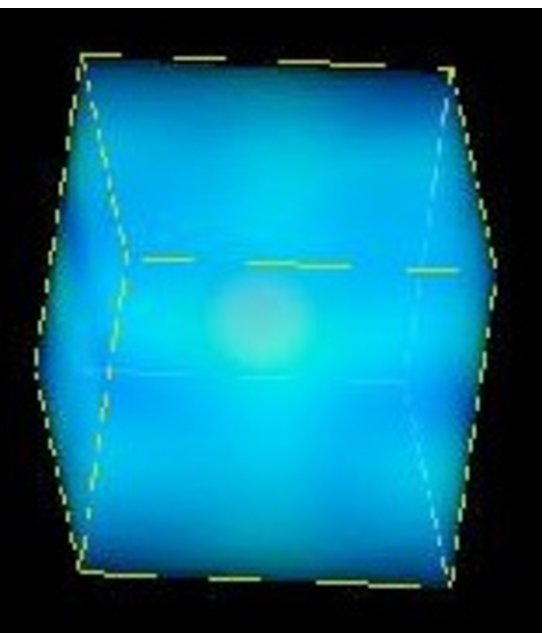}}
 \end{center}
 \caption{Density distributions of nucleons and electron in the case of beta-equilibrium.
 The left panel (a) shows the result reported in Ref.\ \cite{relamaruyama} 
with WS approximation.
The horizontal axis is the distance from the center of the cell, 
and the vertical axis densities of nucleons and electron. 
%
\label{fig beta with WS}}
\end{figure}


\begin{figure}[H]
 \begin{center}
 \subfigure[Binding energy]{
  \includegraphics*[width=0.3\textwidth]{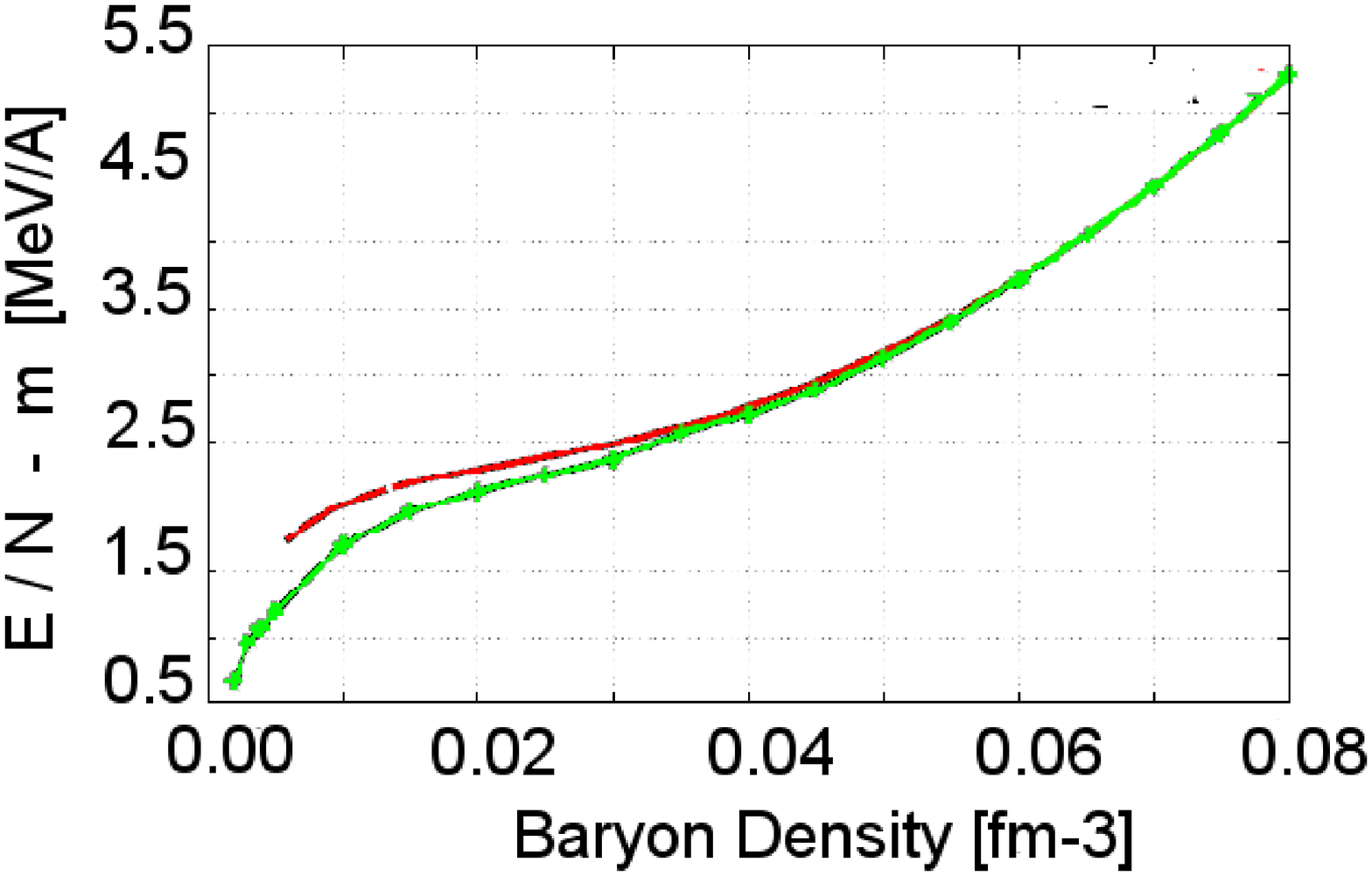}
  \label{Energy beta}}
 \subfigure[Total pressure]{
  \includegraphics*[width=0.3\textwidth]{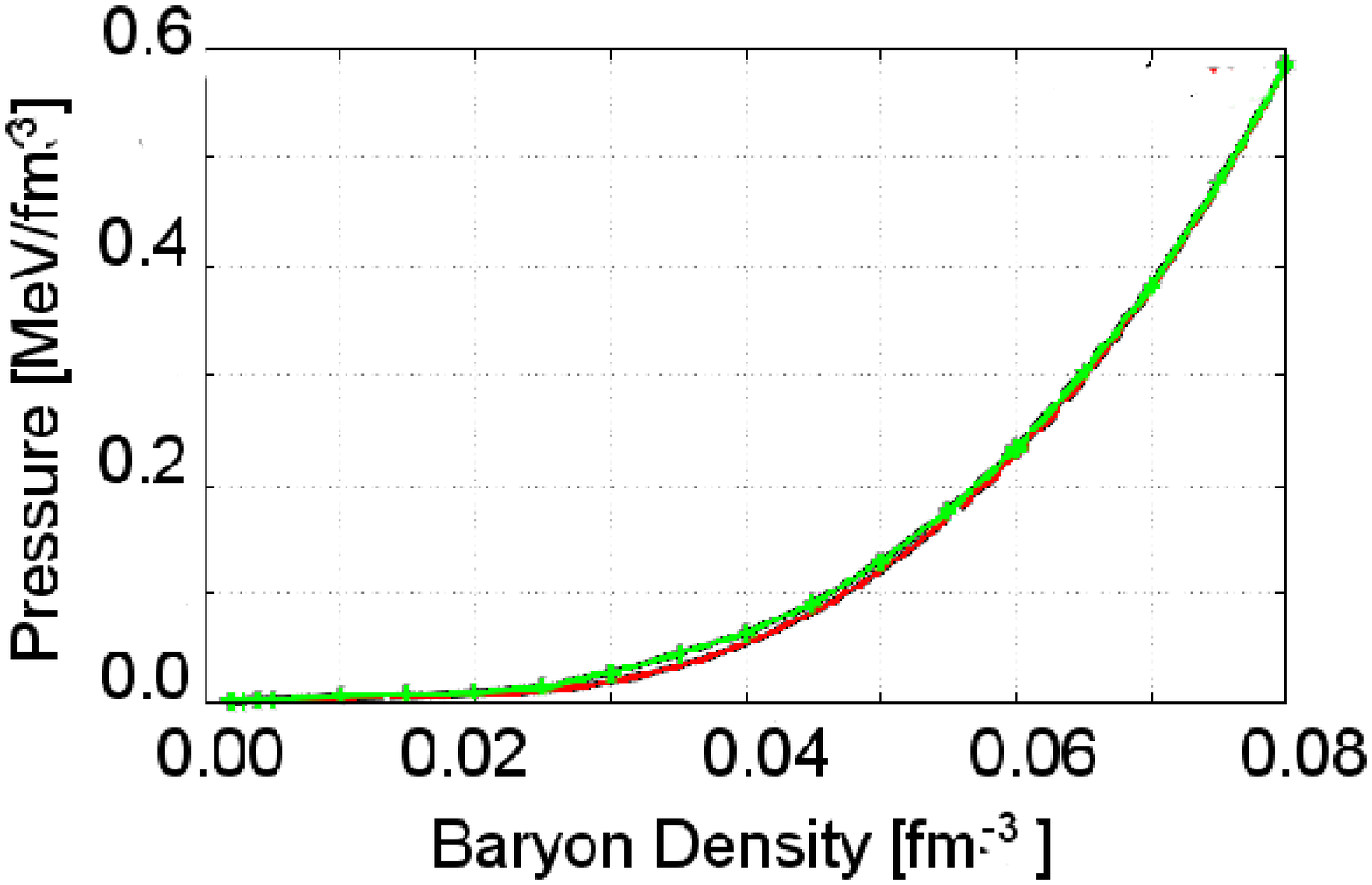}
  \label{Pressure beta}}
 \subfigure[Proton mixing ratio $Y_p$]{
  \includegraphics*[width=0.3\textwidth]{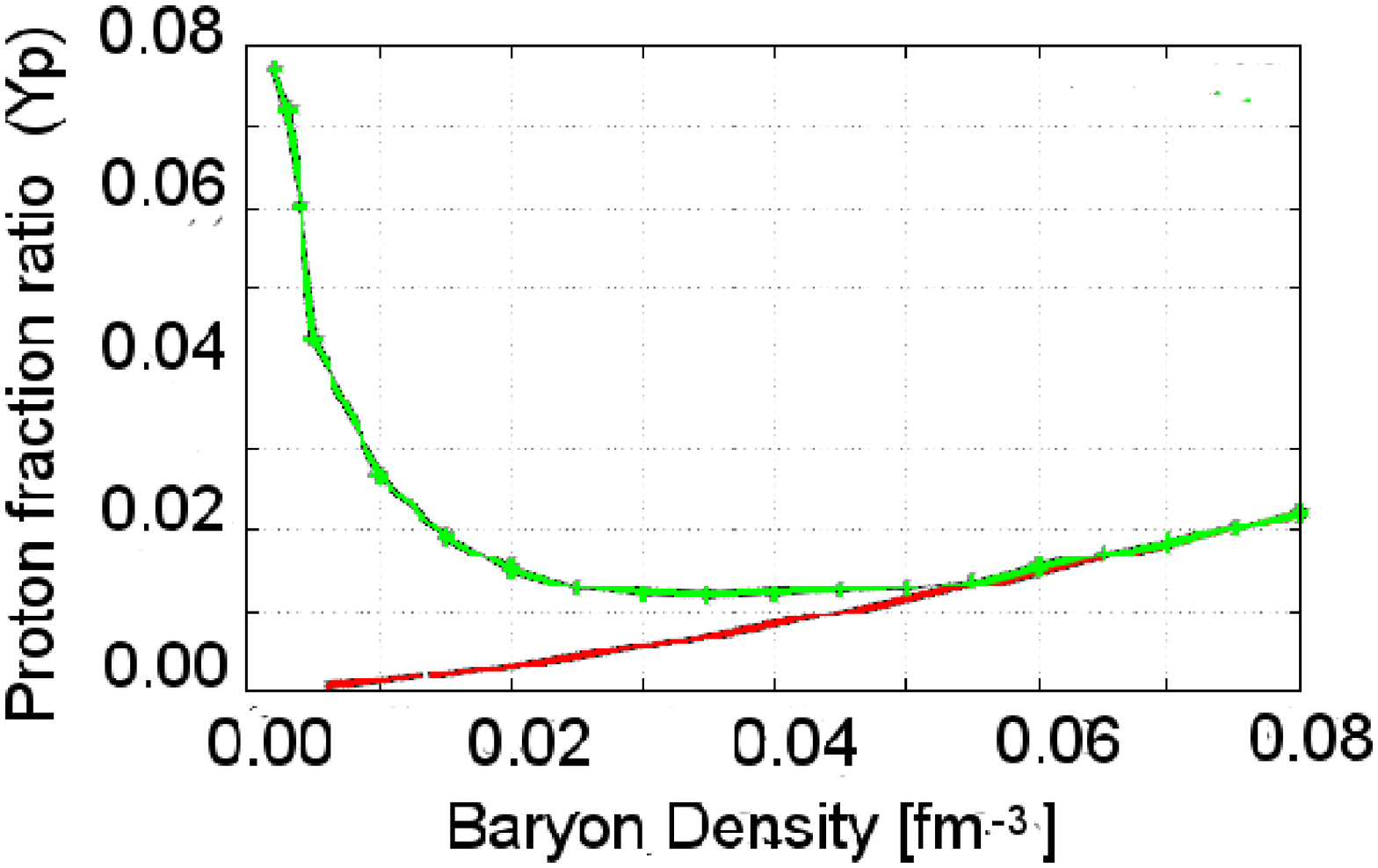}
  \label{Proton mixing ratio beta}}
  \end{center}
 \caption{Energy, pressure, and $Y_p$ of matter in beta-equilibrium. \label{fig beta}}
\end{figure}

In Figs.\ \ref{fig beta with WS} and \ref{fig beta}, we present 
density distributions by the WS cell approximation 
and the present three-dimensional calculation. 
While some pasta structures appeared for fixed proton mixing ratio, 
only the droplet structure appears for the beta-equilibrium case.
This result is qualitatively the same as in the WS cell approximation.

Nuclear matter in the crust region of neutron stars 
includes the very small fraction of protons and electrons. 
So the contribution to the energy and pressure comes 
almost from neutrons and a little from protons and electrons.
Consequently the binding energy and pressure of uniform matter 
are almost the same as those of non-uniform matter.
The proton mixing ratio $Y_p$, however, is significantly different between uniform
and no-uniform cases.
Though it approaches to zero in the zero-density limit for uniform matter, 
there is a steep rise for non-uniform matter.
It approaches to approximately 0.5, which corresponds to 
a neutral atom in the low-density limit.

\section{Conclusion}
We have performed the three-dimensional calculations for low-density nuclear
matter and presented first results, based on relativistic mean-field theory and Thomas-Fermi
approximation.
For some fixed proton mixing ratios and for  $\beta$-equilibrium nuclear matter, 
we have obtained non-uniform structures and the relevant EOS.
The observed structures are typical ``pasta'' structures, 
which were previously studied with the WS cell approximation.
For $Y_p=0.5$, all of the typical pasta structures appeared. 
For $Y_p=0.3$ and 0.1, however, some of the pasta structures 
did not appear.
The density region for each type of pasta structure was also different
between the WS cell approximation and the three-dimensional calculation.

These may be due to the difference of the shapes of the cell.
The WS cell for droplet, rod, slab, tube and bubble structures are
 sphere, cylinder, plate, cylinder and sphere, respectively.
On the other hand, the cubic cell is used in the three-dimensional calculation.
In any way, it is desirable to avoid the effects of the shape of the used cell.
We therefore are planning to enlarge the cubic cell so that several periods
of structures can emerge in it.

For $\beta$-equilibrium, we got only the droplet structure.
This result is qualitatively the same as the one with the WS cell approximation.
However, we cannot definitely state that nuclear matter in $\beta$-equilibrium 
have only droplet, because other works with different interaction have 
observed pasta structures other than droplet.
In $\beta$-equilibrium,
proton mixing ratio is largely different from that in uniform matter.
It tends to zero for uniform matter at zero-density limit,
while it increases in the case of non-uniform matter,
and the asymptotic value should be 0.5 which represents atoms.

\end{document}